\documentclass[11pt]{article}
\pdfoutput=1

\usepackage{jheppub}
\bibstyle{JHEP}

\usepackage[english]{babel} 
\usepackage[utf8]{inputenc}  
\usepackage[T1]{fontenc}
\usepackage{lmodern}
\usepackage{amsmath}         
\usepackage{amsfonts}          
\usepackage{amssymb}
\usepackage{graphicx} 
\usepackage{bbold}
\usepackage{bbm}
\usepackage[colorlinks=true,urlcolor=blue,linkcolor=blue,citecolor=blue,linktocpage=true]{hyperref}

\newcommand{\NN}{{\mathbb{N}}}
\newcommand{\RR}{{\mathbb{R}}}
\newcommand{\ZZ}{{\mathbb{Z}}}
\newcommand{\Tr}{{\mathrm{Tr}}}
\newcommand{\CC}{{\mathcal{C}}}
\newcommand{\CD}{{\mathcal{D}}}
\newcommand{\CF}{{\mathcal{F}}}
\newcommand{\CG}{{\mathcal{G}}}
\newcommand{\CO}{{\mathcal{O}}}
\newcommand{\fg}{{\mathfrak{g}}}
\def\CP{{\cal P}}
\newcommand{\tr}{\mathrm{tr}}
\def\({\left(}
\def\){\right)}
\newcommand{\ket}[1]{| #1 \rangle }
\newcommand{\bra}[1]{\langle #1 |}

\newcommand{\cev}[1]{\reflectbox{\ensuremath{\vec{\reflectbox{\ensuremath{#1}}}}}}

\newcommand{\secref}[1]	{{sec.~\ref{#1}}}

\begin{document} 

\title{Complexity measures from geometric actions on Virasoro and Kac-Moody orbits}

\author{Johanna Erdmenger,}
\author{Marius Gerbershagen}
\author{and Anna-Lena Weigel}

\affiliation{Institute for Theoretical Physics and Astrophysics and W\"urzburg-Dresden Cluster of Excellence ct.qmat, Julius-Maximilians-Universit\"at W\"urzburg, Am Hubland, 97074 W\"{u}rzburg, Germany}

\emailAdd{erdmenger@physik.uni-wuerzburg.de}
\emailAdd{marius.gerbershagen@physik.uni-wuerzburg.de}
\emailAdd{anna-lena.weigel@physik.uni-wuerzburg.de}

\abstract{
  We further advance the study of the notion of computational complexity for 2d CFTs based on a gate set built out of conformal symmetry transformations.
  Previously, it was shown that by choosing a suitable cost function, the resulting complexity functional is equivalent to geometric (group) actions on coadjoint orbits of the Virasoro group, up to a term that originates from the central extension.
  We show that this term can be recovered by modifying the cost function, making the equivalence exact. Moreover, we generalize our approach to Kac-Moody symmetry groups, finding again an exact equivalence between complexity functionals and geometric actions.
  We then determine the optimal circuits for these complexity measures and calculate the corresponding costs for several examples of optimal transformations.
  In the Virasoro case, we find that for all choices of reference state except for the vacuum state, the complexity only measures the cost associated to phase changes, while assigning zero cost to the non-phase changing part of the transformation.
  For Kac-Moody groups in contrast, there do exist non-trivial optimal transformations beyond phase changes that contribute to the complexity, yielding a finite gauge invariant result.
 Furthermore, we also show that the alternative complexity proposal of path integral optimization is equivalent to the Virasoro proposal studied here.
  Finally, we sketch a new proposal for a complexity definition for the Virasoro group that measures the cost associated to non-trivial transformations beyond phase changes.
  This proposal is based on a cost function given by a metric on the Lie group of conformal transformations. The minimization of the corresponding complexity functional is achieved using the Euler-Arnold method yielding the Korteweg-de Vries equation as  equation of motion.
}

\maketitle

\section{Introduction}
An important question in the context of the AdS/CFT correspondence \cite{Maldacena} is how the bulk geometry is encoded in the boundary field theory. In this context, much attention has recently been paid to quantum computational complexity, a concept adapted from quantum information. In general, computational complexity counts the minimum number of simple steps necessary to perform a calculation. In the context of quantum computing, computational complexity is defined as the minimum number of unitary operators, \textit{gates}, necessary to reach a certain target $\ket{\psi_T}$ from a simple reference state $\ket{\psi_R}$. From the point of view of the AdS/CFT correspondence, interest in complexity arose due to similarities between the growth of a black hole interior and time evolution of complexity \cite{Susskind:2014rva}. This observation lead to concrete proposals for holographic complexity in the dual gravity theory \cite{CV, CA}. An open question that remains in view of establishing a concrete AdS/CFT dictionary entry for complexity is how to define it in general quantum field theories, for which the Hilbert space is infinite.
 While a number of complexity definitions have been proposed for free theories (see e.g.~\cite{Jefferson:2017sdb,Chapman:2017rqy,Jiang:2018nzg,Hackl:2018ptj,Khan:2018rzm,Chapman:2018hou, Ge:2019mjt,Guo:2020dsi}, though this list of references is by no means exhaustive), it remains an open question how to define it for interacting and even strongly coupled quantum field theories. For making progress in this direction, it is useful to consider conformal field theories (CFTs) in 1+1 dimensions, for which symmetries impose significant constraints on dynamics and observables.

The foundation of much progress in this direction is the geometric approach that Nielsen formulated for finite-dimensional Hilbert spaces \cite{Nielsen, Nielsen_2, Nielsen_3}. This approach relates the minimum number of gates, i.e. the shortest quantum circuit, to the length of the shortest geodesic in the space of unitary operators. Instead of a discrete chain of gates acting on the reference state, this approach is based on a path ordered exponential of a time-dependent Hamiltonian referred to as the instantaneous gate $Q(t)$,
\begin{equation}
  U(T) = \cev{\CP}\, \mathrm{exp} \left[-\int_0^T Q(t)dt\right].
  \label{eq:unitary transformation}
\end{equation}
The simple counting of gates is replaced by a notion of cost associated to the path in the space of unitaries,
\begin{equation}
  \CD(U(T)) = \int_0^T \! dt \, \CF(U(t),\dot U(t)).
\end{equation}
Here, $\CF(U(t),\dot U(t))$ is the \textit{cost function}, a functional in the space of unitaries that tells us how expensive the application of the infinitesimal unitary operator along the path at time $t$ is.
The computational complexity is then obtained by minimizing $\CD(U(T))$ with respect to all possible paths in the space of unitaries taking us from the reference state $\ket{\psi_R}$ to the target state $\ket{\psi_T}$.

Based on Nielsen's approach, a recent proposal by Caputa and Magán \cite{Magan} for defining complexity in quantum field theories suggests to restrict the allowed set of gates to symmetry transformations.
Then, the complexity may be computed in terms of compositions of infinitesimal symmetry transformations, which represent the gates.
A particular advantage of this method is that the cost function $\CF$ is fixed up to a prefactor and a choice of norm.
This proposal was employed to compute complexity for conformal transformations in 2d CFTs \cite{Caputa:2018kdj}.
The cost functions used in \cite{Magan,Caputa:2018kdj} are state dependent, in contrast to the approach of Nielsen where the cost and thus the complexity depends only on the path taken through the space of unitaries.
As a consequence of this choice, the expression for the cost $\CD(U(T))$ proposed in \cite{Caputa:2018kdj} takes the form of an action functional which is dependent on the reference state and on the path through the space of unitaries.
Equivalently, because only symmetry transformations are considered, the action is dependent on a path through the symmetry group manifold.
Complexity is defined as the minimum of this action functional, i.e.~as the on-shell value after variation w.r.t.~the group elements specifying the path from the reference to the target state.
Strikingly, in \cite{Caputa:2018kdj}  the action functional obtained in this way was found to be equivalent to two different actions.
The first one of these is the well-known Polyakov action for induced two-dimensional gravity \cite{Polyakov:1987zb}.
Moreover, it was found that the complexity functional is equivalent to a \textit{geometric action} defined on a coadjoint orbit \cite{Kirillov, Witten, Alekseev, Alekseev_3, Alekseev_4, Barnich} of the Virasoro group.
Geometric actions in general may be thought of as an analogue of Hamiltonian actions of classical systems defined on phase spaces, where the phase space is now a coadjoint orbit.
In \cite{Caputa:2018kdj},  terms arising from the central extension of the Virasoro group were discarded from the on-shell action functional in establishing this equivalence. These terms were argued not to spoil the equivalence in the large central charge limit. As we discuss in the present paper however, these terms do play an important role.

This publication is dedicated to the further study of the proposal of \cite{Magan,Caputa:2018kdj} and in particular of  its relation to geometric actions.
We also study the relation between the proposal of \cite{Magan,Caputa:2018kdj}
and an alternative complexity proposal based on 
 path integral optimization  \cite{Caputa:2017urj}.
This proposal, at first glance completely different, is based on intuition from tensor networks.
Its starting point is the fact that ground-state wave functionals may be computed by Euclidean path integrals.
These path integrals may be discretized by using a lattice regularization.
The geometry of this lattice regularization is not fixed and can be changed by modifying the background metric for the path integration without introducing a change in the wave functional.
The authors of \cite{Caputa:2017urj} then argue that the number of lattice points is a measure of the complexity needed to prepare the state whose wave functional is computed by the path integral.
For 2d CFTs, under a change of the background metric, the path integral changes by an on-shell value of the Liouville action. In \cite{Caputa:2017urj},
the minimum of this on-shell value is identified with the complexity.
The minimization is taken over all possible background metrics, subject to boundary conditions that fix the path integral to the wave functional for the target state in question.
In this way, background metrics that are time slices of asymptotically AdS$_3$ spacetimes are obtained, for which the complexity is equal to the volume of the corresponding time slice.
See also \cite{Caputa:2017yrh,Bhattacharyya:2018wym,Takayanagi:2018pml} for further developments in this direction.

Before we turn to the main questions addressed in the present paper, let us briefly mention
further related work on geometric actions and complexity. This includes \cite{Bueno:2019ajd,Akal:2019hxa,Ghodrati:2019bzz}.
\cite{Bueno:2019ajd} studied in a general setup the conditions under which cost functions provide lower bounds to the circuit complexity for a discrete set of quantum gates.
The authors find that lower bounds are best achieved by considering cost functions whose associated complexity functionals are closely related to geometric actions.
In \cite{Akal:2019hxa}, connections were found between the complexity definition of \cite{Caputa:2018kdj} and Berry phases in unitary representations of the Virasoro group that have recently been studied in \cite{Berry_phases}.
In \cite{Ghodrati:2019bzz}, the path integral optimization procedure of \cite{Caputa:2017urj} is generalized to the warped AdS$_3$/warped CFT$_2$ correspondence.
Moreover, \cite{Ghodrati:2019bzz} also considers the application of the complexity proposal of \cite{Magan,Caputa:2018kdj} to the case of Kac-Moody symmetries -- a point that we will study in more detail in the present paper.

In this paper, we aim at answering three questions:  Can the cost function proposed in \cite{Caputa:2018kdj} be modified in such a way that the complexity of symmetry transformations in CFTs becomes exactly equal to the geometric action, including the central extension terms discarded previously? In this case, the complexity will simply be given by the on-shell value of the geometric action. This will hold the advantage that we may use the well-understood geometric actions to gain valuable insight into CFT complexity. The properties of the geometric action, such as gauge invariance under certain subgroups called stabilizers, will directly influence the complexity measurement. This leads us to the second question: Do geometric actions provide good complexity measures, in the sense that they are physically meaningful? Here again, the properties of  geometric actions provide  valuable means to address this question. And finally, the third question is: What is the relation between this notion of complexity and the path integral approach proposed in~\cite{Caputa:2017urj}?

For answering the first question, we obtain a precise relation between the complexity and the geometric action by identifying corresponding elements in both actions.
In particular, we perform this identification not only for the Virasoro group, but also for  generalizations to CFTs with Kac-Moody symmetries. 
In addition to the usual conformal symmetries, these CFTs have a symmetry given by a semisimple Lie group.
We show that the complexity functional obtained for Kac-Moody symmetries shares the same similarities with the geometric action already observed in \cite{Caputa:2018kdj} for conformal symmetries: The complexity is equal to the geometric action up to terms arising from the central extension of the corresponding symmetry group.
We then proceed to deriving a new generalized cost function that is sensitive to contributions from the central extension of the symmetry group\footnote{Geometric actions are in general defined only up to addition of a Hamiltonian term $\int dt H_X$, where $H_X$ is a Hamiltonian function on the coadjoint orbit \cite{Barnich,Cotler:2018zff}. The additional contribution to the cost function that we consider here is not of this form: It is a term arising from the central extension that we show to be necessary to obtain the geometric action. In the following, we choose a vanishing Hamiltonian $H_X=0$ on the coadjoint orbit.}.
With this new cost function, the complexity then becomes exactly equal to the geometric action of the considered symmetry group (Kac-Moody or Virasoro).
This is one of the main results of this work.

We note an aspect that distinguishes our work from the complexity proposal by Caputa and Magán \cite{Caputa:2018kdj}.
A new element of the present paper is that in contrast to the definition of \cite{Caputa:2018kdj}, we consider an additional contribution to the cost function that involves the central extension of the Maurer-Cartan form.
While it was argued previously that this central extension term does not lead to additional contributions to the equations of motion, we show that these terms are indeed physically significant.
In particular, we find that the equations of motion for both cost functions are different\footnote{See \eqref{eq:eom_Virasoro} and \eqref{eq:eom_complexity_Caputa_Magan}.} and only when we include the central extension, the complexity functional becomes equivalent to a geometric action.
Hence, the additional contribution has to  be taken into account when deriving equivalence statements between geometric actions and complexity functionals.

In order to answer the second question whether geometric actions provide viable complexity measures, we determine the optimal transformations, i.e.~those with minimal cost, for both the Virasoro and Kac-Moody groups.
We then use these results to compute the  complexity for simple examples of optimal transformations.
We demonstrate that for both symmetry groups, the complexity is non-vanishing for transformations changing only the phase of the reference state.
Moreover, for the Virasoro group, we obtain different costs for identical transformations.
We explain that this inconsistency is caused by a lack of gauge invariance of the geometric action under transformations that relate physically indistinguishable states.
Generally, invariance of actions implies invariance only up to boundary terms that leave the equations of motion invariant.
Since we identify the on-shell value of the geometric action with the complexity, these boundary terms contribute to the complexity and thus must be added to the geometric action to obtain consistent results. 
Adding these terms cancels the cost of phase changes.
Moreover, we observe that a general optimal transformation for the Virasoro group instantaneously jumps to the target state up to phase.
For all reference states except for the vacuum, the complexity assigns zero cost to the instantaneous jump, hence yielding zero once the boundary term is added to the action.

A further interesting observation is that upon adding the boundary terms, the Virasoro complexity becomes a special case of a Virasoro Berry phase, as introduced in \cite{Berry_phases}. This Berry phase arises from unitary highest-weight representations of the Virasoro group when certain conformal transformations are applied to a primary state. These transformations must form a closed path under projection onto the coadjoint orbit associated with the state. The equivalence of our Virasoro complexity functional and the Berry phases considered in \cite{Berry_phases} further illustrates that we cannot expect the Virasoro complexity to count anything but a phase change. Moreover, the Virasoro complexity is even more restrictive than the Berry phase as the latter only requires the path on the coadjoint orbit to be closed, whereas the complexity functional additionally demands the transformation be optimal. 

In contrast, the space of optimal transformations obtained from the Kac-Moody complexity functional is much larger: An optimal circuit is given by the product of two matrices that belong to the specific semisimple group considered. It only has to satisfy certain mild conditions. Therefore, there exist many non-trivial optimal transformations for the Kac-Moody group. Non-trivial transformations map the reference state, which we take to be a highest weight state, to a particular linear combination of its descendants. Trivial transformations, on the other hand, change only the phase of the reference state.
We give examples of non-trivial transformations for several Kac-Moody groups and show that their associated complexity is non-vanishing.
While as in the Virasoro case, additional boundary terms are required to ensure gauge invariance of the action under the orbit stabilizer group, the complexity of optimal paths is generally non-zero even when these terms are added. This implies that in contrast to the Virasoro case, for Kac-Moody groups geometric actions do provide physically viable complexity measures.

We then move on to answer the third question and show how the complexity resulting from geometric actions is related to the path integral approach of \cite{Caputa:2017urj}. We achieve this by identifying solutions of the Liouville equations leading to path integral complexity with solutions of the Virasoro geometric action. In this way we show how the path integral optimization approach of \cite{Caputa:2018kdj} is related to more conventional notions of complexity based on gate sets as well as on target and reference states\footnote{Note that subtracting the aforementioned boundary terms from the Virasoro complexity spoils the equivalence between both approaches.}.
We then show that the examples considered in \cite{Caputa:2017urj} in the context of path integral complexity correspond to transformations again leading just to a phase change in the geometric action approach.
This suggests that the complexity definition based on geometric actions as considered in the present paper does not provide a physically meaningful interpretation of the path integral approach.
This does not, however, invalidate the path integral approach, but simply implies that one should look elsewhere for its  physical interpretation.

Our results show that for the geometric actions, in many cases (and in particular for the Virasoro group)  the optimal paths jump instantaneously to the target state with the complexity definition measuring only trivial phase changes.
To obtain a non-trivial complexity definition for conformal transformations, we propose a different approach based on an idea put forward in \cite{Caputa:2018kdj}. This proposal is based on the Euler-Arnold equations describing the geodesic flow on a group manifold.
Equipping the tangent space of this manifold with a metric, the complexity may then be determined as the length of a geodesic in the group manifold and hence in terms of the length of the velocity vector.
For the choice of metric given by the inner product of the Virasoro algebra, the equations of motion are given by the well-known Korteweg-de Vries equation.
We show that the space of solutions of this equation, i.e.~the space of optimal transformations, is much larger than that of the geometric action.
In particular, we demonstrate that, in contrast to the geometric action approach, many non-trivial transformations exist that do not jump instantaneously to the target state.
Although we do not determine the most general optimal transformations for this approach explicitly, the existence of these non-trivial transformations shows that the approach is a promising proposal for defining a non-trivial complexity for conformal transformations.

The outline of our paper is as follows.
In \secref{sec:complexity_general}, we review the complexity proposal \cite{Caputa:2018kdj} of Caputa and Magán and generalize this approach to the case of two dimensional CFTs with Kac-Moody symmetry groups.
In \secref{sec:geometric actions}, we present a short introduction to coadjoint orbits and geometric actions.
This mathematical framework is then used in \secref{sec:complexity and geometric actions} to show that by a modification of the approach of \cite{Caputa:2018kdj} the equivalence between geometric actions and the complexity action functionals becomes exact for both the Virasoro and Kac-Moody group.
Sec.~\ref{sec:Examples} is dedicated to the study of optimal transformations and their associated complexity.
In \secref{sec:Gravity and Liouville theory}, we briefly comment on coadjoint orbit actions arising from 3d gravity and their equivalence to the complexity expression derived before.
Moreover, we find a precise connection between the complexity definitions of \cite{Caputa:2018kdj} and \cite{Caputa:2017urj}.
To address the issues encountered in \secref{sec:Examples},  we sketch a modified complexity proposal based on a different cost function arising from the Euler-Arnold formalism in \secref{sec:Euler-Arnold complexity}.
Finally, we present our conclusions in \secref{sec:discuss}.

\section{Complexity for Kac-Moody and Virasoro groups}
\label{sec:complexity_general}

	In this section, we review how complexity may be obtained for $2$d CFTs. Since these calculations concern infinite-dimensional systems, they necessarily differ in certain aspects from Nielsen's geometric approach to finite-dimensional problems. We discuss where both approaches deviate and why it is useful to restrict the allowed set of gates to symmetry transformations as suggested in \cite{Magan}. The general procedure to obtain complexity for $2$d CFTs is described in sec.~\ref{sec:complexity_symmetry_group}. We then review its application to the Virasoro group based on \cite{Caputa:2018kdj} in sec.~\ref{sec:Virasoro_complexity} and compute complexity for Kac-Moody groups in sec.~\ref{sec:complexity_Kac-Moody}. 

\subsection{Complexity for symmetry groups}
\label{sec:complexity_symmetry_group}
Since Nielsen's geometric approach relates a discrete gate-counting procedure to finding geodesics in a smooth manifold, it may be applied to continuous systems. Nevertheless, as discussed in \cite{Magan}, in order to obtain complexity for CFTs some modifications are necessary. 

First, Nielsen's cost functions $\CF$ involve penalty factors. As the name suggests, these penalize certain unitary gates that are associated with difficult transformations and are thus expected to be more costly. Unfortunately, a prescription for choosing these penalty factors in order to match holographic complexity proposals does not exist, introducing some arbitrariness into complexity calculations. This was one of the reasons in \cite{Magan} to restrict the allowed set of gates to symmetry transformations. If the system of interest is invariant under these transformations, all infinitesimal symmetry transformations should be equally difficult to apply. The penalty factors then reduce to an overall prefactor, bearing no relevance on the form of the complexity functional itself. Of course, by restricting the set of allowed gates to symmetry transformations,  the set of possible target states is restricted to those reachable by symmetry transformations. 

Let $\ket{\psi_T}$ be a possible target state. The reference state $\ket{\psi_R}$ may be conveniently chosen. Consequently, the complexity of the target state always depends on the particular choice of reference state. Let $G$ be a symmetry group of the system of interest. Unitary representations of group elements $g$ are denoted by $U_g$. Then, the circuit connecting the reference and target state is a path through the unitary manifold, and the target state is given by 
\begin{equation}
\ket{\psi_T} =U_{g(T)}\ket{\psi_R},
\end{equation} 
where $T$ specifies the time after which the target state is reached. Additionally, we require $U_{g(t=0)}=1$ to ensure the initial state is the reference state. In accordance with Nielsen's geometric approach, the complexity of the target state is given by the length of the shortest path connecting the reference and the target state in the unitary manifold. The finite transformation $U_{g(T)}$ may be decomposed into infinitesimal transformations, $U_{g(T)}=U_{\epsilon(T)}U_{\epsilon(T-dt)}...U_{\epsilon(dt)}1$. Since we are considering symmetry groups, the gates are thus infinitesimal symmetry transformations and may be written in terms of the generators. 

Let $J$ denote the conserved current and $\epsilon$ the velocity, specifying which generator is applied at a given time. Then, the gate reads
\begin{equation}
Q(t)=\frac{1}{2\pi}\int dx\, \epsilon(t,x)J(x).
\label{eq:gate}
\end{equation}
Note that the gate \eqref{eq:gate} is path ordered according to \eqref{eq:unitary transformation}, which implies that earlier gates are applied first. In other words, the time-dependent velocity $\epsilon(t,\sigma)$ specifies which infinitesimal symmetry transformation is applied at the given time.
Infinitesimally close points along the path are related by,
\begin{equation}
U(t+dt)=e^{-Q(t)dt}U(t),
\end{equation}
where $U_{g(T)}=U(T)$.  
The instantaneous velocities $\epsilon$ may be computed in terms of group elements $g$ by employing
\begin{equation}
g(t+dt, x)=e^{\epsilon(t,x)dt}g(t,x)
\label{eq:velocity}
\end{equation}
and expanding to first order.  

Finally, we need to specify an appropriate cost function, which assigns computational cost to every symmetry transformation. At this point, we return to the point that Nielsen's approach requires  modifications in the context of CFTs. While the issues caused by unknown penalty factors are readily solved by restricting the gates to symmetry transformations, we have yet to address the effects of infinite-dimensional manifolds as encountered in CFTs.

Nielsen considered finite-dimensional systems with cost functions that do not yield finite results in infinite-dimensional systems. A remedy is obtained by introducing an explicit state-dependence in the cost function, i.e. the gate is evaluated in the state present at any given time $t$ along the path. This corresponds to introducing a density matrix $\rho(t)=U(t)\rho_0 U^{\dagger}(t)$, where the transformations $U(t)$ evolve the initial density matrix $\rho_0=\ket{\psi_R}\bra{\psi_R}$. Then, the state-dependent one-norm cost function\footnote{At this point, it is worth mentioning that the choice of cost function is not unique. For example, a closely related possibility is the state-dependent two-norm. For a recent discussion of possible cost functions and their respective merits, we refer to \cite{Bueno:2019ajd}. We choose the one-norm since the complexity functional obtained with this cost function yields the geometric action of the corresponding symmetry group up to terms arising from the central extension of the group. This will later allow us to obtain the full geometric action by slightly modifying this cost function. } is given by 
\begin{equation}
\CF=|\tr[\rho(t)Q(t)]|=|\bra{\psi_R }U^{\dagger}(t)Q(t)U(t)\ket{\psi_R}|.
\label{eq:1_norm_cost_function}
\end{equation}
The total cost of a possible path connecting the reference and target state is then given by 
\begin{equation}
\CC=\int dt\,\CF=\frac{1}{2\pi}\int dt\int dx\,\epsilon(t,x)\bra{\psi_R}U^{\dagger}(t)J(x)U(t)\ket{\psi_R}.
\label{eq:complexity_general_CM}
\end{equation}
The complexity, i.e.~the minimal cost, is obtained from \eqref{eq:complexity_general_CM} as follows. First, find expressions for the velocity $\epsilon(t,\sigma)$ and the transformed current $U^{\dagger}(t)J(x)U(t)$ in terms of the group paths $g(t)$. Then, choose a reference state and evaluate the expectation value in \eqref{eq:complexity_general_CM}. 
Finally, to minimize the new expression for \eqref{eq:complexity_general_CM}  given in terms of the group path $g(t)$, solve the equation of motion arising from \eqref{eq:complexity_general_CM}. The solutions correspond to optimal circuits. Choose the solution that yields the desired target state upon application to the chosen reference state. Inserting this solution into the complexity measure \eqref{eq:complexity_general_CM} gives rise to its appropriate on-shell value. This on-shell value is the complexity associated with the optimal transformation that yields the desired target state.
-- In \secref{sec:Examples} this procedure is applied to the modified complexity functional introduced in \secref{sec:complexity and geometric actions}.

We now review the application of the above method to the Virasoro group based on \cite{Caputa:2018kdj} in sec.~\ref{sec:Virasoro_complexity} and present new results for the Kac-Moody group in sec.~\ref{sec:complexity_Kac-Moody}.

\subsection{Virasoro group}
\label{sec:Virasoro_complexity}
Before we proceed with the Kac-Moody group, we now briefly review the results of \cite{Caputa:2018kdj} on applying the procedure of the previous subsection to the Virasoro group.

The Virasoro group is centrally extended, and we denoted group elements by $(f(\sigma), \alpha)$, where $\sigma\in S^1$. The first element $f(\sigma)$ are the orientation-preserving diffeomorphisms of the circle $\mathrm{Diff}^+( S^1)$. In more physical terms, $f(\sigma)$ are conformal transformations in two dimensions; for example, upon interpreting $\sigma$ as a light-cone coordinate $x^{+}$, $f\in \mathrm{Diff}^+( S^1)$ are coordinate transformations $x^{+}\rightarrow f(x^{+})$. Since the group is centrally extended, there is an additional number $\alpha\in \mathbb{R}$ from the central extension of the group. The cost function \eqref{eq:complexity_general_CM} does not associate any cost to contributions from the central extension. We may hence disregard $\alpha$ for the moment and consider only the conformal transformations $f(t, \sigma)$. We comment on the significance of the central extension for deriving the equivalence of the complexity functional to a geometric action later on in \secref{sec:complexity and geometric actions}.

To every diffeomorphism $f$, there exists an inverse diffeomorphism $F$ such that at any given time $f(t, F(t, \sigma))=\sigma$. The gate \eqref{eq:gate} may be written in terms of the conserved energy-momentum tensor $T$,
\begin{equation}
Q(t)=\frac{1}{2\pi}\int d\sigma \epsilon(t,\sigma)T(\sigma).
\end{equation}
The velocities $\epsilon$ may be computed from the gate equation \eqref{eq:velocity} in terms of the group path $f(t, \sigma)$. Since the resulting expression involves both the diffeomorphism $f$ and its inverse $F$, it proves convenient to recast the velocity in terms of $F$ only, 
\begin{equation}
\epsilon(t, \sigma)=-\frac{\dot{F}(t, \sigma)}{F^{\prime}(t, \sigma)}.
\label{eq:velocity_Vir}
\end{equation}
To calculate the complexity functional \eqref{eq:complexity_general_CM}, we still need an expression for the transformation of the energy-momentum tensor under conformal transformations. Since the velocity \eqref{eq:velocity_Vir} is given in terms of the inverse $F$ rather than the diffeomorphism $f$ itself, it is convenient to write the transformation law as a function of $F$, too. It is given by
\begin{equation}
U^{\dagger}_fTU_f=F^{\prime\,2}T(F)+\frac{c}{12}\{F, \sigma\},
\label{eq:transformation_T}
\end{equation} 
where $\{F,\sigma\}$ is the Schwarzian derivative,
\begin{equation}
\{F,\sigma\}=\frac{F^{\prime\prime\prime}(\sigma)}{F^{\prime}(\sigma)}-\frac{3}{2}\(\frac{F^{\prime\prime}(\sigma)}{F^{\prime}(\sigma)}\)^2.
\label{eq:Schwarzian}
\end{equation}
A convenient reference state is the highest-weight state $\ket{h}$. With the results \eqref{eq:velocity_Vir} and \eqref{eq:transformation_T}, the complexity \eqref{eq:complexity_general_CM} is then given by 
\begin{equation}
\CC=\int dt\,\CF=\frac{1}{2\pi}\int dt\int d\sigma\left[-\dot{F}F^{\prime}\bra{h}T(F)\ket{h}-\frac{c}{12}\frac{\dot{F}}{F^{\prime}}\{F, \sigma\}\right].
\label{eq:complexity_Vir}
\end{equation}
Whereas this expression is useful in the context of geometric actions on coadjoint orbits, which will be the focus of the next section, in order to explicitly calculate the complexity, it is more convenient to recast the functional in terms of the group path $f$,
\begin{equation}
\CC[f] =\frac{1}{2\pi}\int_{0}^{T}dt\int_{0}^{2\pi}  d\sigma\,\frac{\dot{f}}{f^\prime}\left(-\bra{h}T(\sigma)\ket{h} + \frac{c}{12}\{f, \sigma\}\right).
\label{eq:complexity_Vir_intermediate}
\end{equation}
Finally, the expectation value may be evaluated using the mode expansion on the cylinder,
\begin{equation}
T(\sigma)= \sum_n \(L_n-\frac{c}{24}\delta_{n,0}\)e^{-in\sigma}.
\label{eq:T_cyl}
\end{equation}
Here, the zero-mode shift $L_0\rightarrow L_0-\frac{c}{24}$ is a consequence of a change in the ground-state energy on the cylinder with respect to the plane.
Inserting \eqref{eq:T_cyl} into the complexity functional \eqref{eq:complexity_Vir_intermediate} gives
\begin{equation}
\CC[f] =\frac{c}{24\pi}\int_{0}^{T}dt\int_{0}^{2\pi}  d\sigma\,\frac{\dot{f}}{f^\prime}\left( \frac{1}{2}\(1-\frac{24h}{c}\)+\{f, \sigma\}\right).
\label{eq:complexity_CM_Vir}
\end{equation}
In particular, only $L_0$ contributes upon evaluating the expectation value since the highest-weight state $\ket{h}$ is annihilated by all $L_n$ with $n>0$, and the generators $L_{-n}$ with $n>0$ yield states $\ket{h+n}$ satisfying the orthogonality condition $\langle h|h+n\rangle=\delta_{n,0}$. The eigenvalue of $L_0$ is denoted by $h$.\\
An analogous expression to \eqref{eq:complexity_CM_Vir} may be obtained for antiholomorphic transformations.

\subsection{Kac-Moody group}
\label{sec:complexity_Kac-Moody}
We now proceed to consider a CFT with an additional symmetry given by a semisimple Lie group $G$. Let $T^{a}_\lambda$ denote the generators of $G$ in a representation $\lambda$. The finite Lie algebra may be generalized to an infinite-dimensional algebra on a circle in such a manner that it contains the original algebra as a subalgebra, and the Virasoro algebra is a subalgebra of its enveloping algebra. The resulting algebra is a Kac-Moody algebra, with an associated Kac-Moody group.  Note that there exist many different Kac-Moody groups. They are classified by the semisimple Lie group on which they are based. The relation to the Virasoro group implies that the Kac-Moody group is centrally extended. The analogue of the Virasoro central charge $c$ is the level $k\in \mathbb{Z}$ in the Kac-Moody case. Since we aim at computing the complexity functional for CFTs with Kac-Moody symmetries in analogy to sec.~\ref{sec:Virasoro_complexity}, we need the following ingredients: the conserved current for the gate \eqref{eq:gate}, the transformation of the current under symmetry transformations as required by \eqref{eq:1_norm_cost_function} and an appropriate reference state. 

Let us begin with the currents and their transformation. We denote the conserved currents of the Kac-Moody symmetry by $J$. Consequently, the gate \eqref{eq:gate} reads
\begin{equation}
Q(t)=\frac{1}{2\pi}\int d\sigma\, \epsilon(t,\sigma)J(\sigma).
\end{equation}
The currents $J$ may be expanded in terms of the generators $T^{a}_\lambda$ of the semisimple Lie group $G$,  
\begin{equation}
J=\sum_{n\in \mathbb{Z}}\sum_{a=1}^{\mathrm{dim}\,\mathfrak{g}} J_n^a T^a_{\lambda}.
\label{eq:Kac-Moody_mode_expansion}
\end{equation}
Here, $J^a_n$ are the generators of the infinite-dimensional Kac-Moody algebra. 

Next, we derive how the currents transform under symmetry transformations $\Omega\in G$.  This is accomplished by considering Kac-Moody group elements, which are denoted by $(g, \alpha)$. Just as the Virasoro group, the Kac-Moody group is centrally extended, and thus $\alpha\in\mathbb{R}$ belongs to the center, while $g(\sigma)$ are transformations valued in $G$. In particular, $g$ now is a map $g: S^1\rightarrow G$, i.e. we consider maps from the unit circle into the group manifold of the semisimple group $G$. In terms of the Kac-Moody group element $g$, the conserved current associated with left-multiplication symmetry $g \rightarrow \Omega g$ is given by $J=-k\partial_\sigma gg^{-1}$. The transformation of the currents under symmetry transformations $\Omega$ then directly follows from the transformation property of $g$,
\begin{equation}
U^{\dagger} JU=\Omega(t, \sigma) J(\sigma)\Omega^{-1}(t,\sigma)-k\Omega^{\prime}(t, \sigma)\Omega^{-1}(t, \sigma).
\label{eq:transformation_current}
\end{equation}
A similar expression may be obtained for the antiholomorphic current.

Next, we derive the velocity $\epsilon$ from the gate equation \eqref{eq:velocity}, which for a group path $\Omega$ in $G$ reads
\begin{equation}
\Omega(t+dt, \sigma)=e^{\epsilon(t,\sigma)dt}\Omega(t, \sigma).
\end{equation}
Expanding to first order, we obtain the velocity in terms of the transformations $\Omega$,
\begin{equation}
\epsilon(t, \sigma)=\dot{\Omega}(t, \sigma)\Omega^{-1}(t, \sigma).
\label{eq:velocity_Kac-Moody}
\end{equation}

Finally, we pick an appropriate reference state. Just as in sec.~\ref{sec:Virasoro_complexity}, a highest-weight state $\ket{\Phi}$ is most convenient. In particular, $\ket{\Phi}$ is a highest-weight state in the same sense as $\ket{h}$ for the Virasoro group, i.e. it is annihilated by all $J^{a}_n$ with $n>0$. Moreover, $J^a_{-n}$ with $n>0$ yield states orthogonal to $\ket{\Phi}$.

We now have all necessary ingredients to obtain an expression for the complexity. With \eqref{eq:velocity_Kac-Moody} and \eqref{eq:transformation_current}, \eqref{eq:complexity_general_CM} becomes
\begin{equation}
\CC=\int dt\,\CF=\frac{1}{2\pi}\int dt\int d\sigma\,\tr\left[\dot{\Omega}\Omega^{-1}\bra{\Phi}J(\sigma)\ket{\Phi}-k\dot{\Omega}\Omega^{-1}\Omega^{\prime}\Omega^{-1}\right].
\label{eq:complexity_Kac-Moody_CM}
\end{equation}
To evaluate the expectation value, we use the mode expansion \eqref{eq:Kac-Moody_mode_expansion} and the action of the generators on the highest-weight state $\ket{\Phi}$ as described above. Since the action of the generators is completely analogous to the Virasoro case, we observe here as well that only $J_0$ survives the expectation value. Its eigenvalue is the generator $-T^{a}_\lambda$ in the representation $\lambda$.
Thus, the complexity functional \eqref{eq:complexity_general_CM} takes the form 
\begin{equation}
\CC=\frac{1}{2\pi}\int_0^Tdt\int_0^{2\pi} d\sigma\, \tr\Big[-\dot{\Omega}\Omega^{-1}\sum_{a=1}^{\mathrm{dim}\,\mathfrak{g}} T^a_\lambda T^a_\lambda-k\dot{\Omega}\Omega^{-1} \Omega^{\prime} \Omega^{-1}\Big],
\label{eq:complexity_KM}
\end{equation}
where $\sum_{a=1}^{\mathrm{dim}\,\mathfrak{g}}T^a_\lambda T^a_\lambda$ is the Casimir element $C_{\lambda}$ in the representation $\lambda$.
This is our final result for the generalization of the complexity definition of \cite{Caputa:2018kdj} to the case of a Kac-Moody symmetry group.
In \secref{sec:complexity and geometric actions}, we comment on its interpretation and in particular the relation to geometric actions, a topic which we now introduce in \secref{sec:geometric actions}.

\section{Geometric actions on coadjoint orbits}
\label{sec:geometric actions}
It was observed in \cite{Caputa:2018kdj} that the complexity functional \eqref{eq:complexity_CM_Vir} is closely related to the geometric action on a coadjoint orbit of the Virasoro group.
In particular, \eqref{eq:complexity_CM_Vir} is equal to this geometric action up to terms which we show in \secref{sec:complexity and geometric actions} to be arising from the central extension of the Virasoro group.
To this end, we now give a brief introduction to geometric actions.
Sec.~\ref{sec:geometric_action} contains the basic definitions for the general case.
Sec.~\ref{sec:central_extensions} introduces central extensions, for which geometric actions are constructed in \secref{sec:geo_action_with_central_extension}.
Finally in \secref{sec:geo_Vir_Kac} we present geometric actions for Virasoro and Kac-Moody symmetry groups.
  
\subsection{Coadjoint orbits as symplectic manifolds }
\label{sec:geometric_action}
Geometric actions were first introduced by Kirillov \cite{Kirillov} in the context of geometric quantization. Here, we briefly introduce relevant concepts to construct the geometric action.

Let $G$ be a Lie group with group elements $g$ and Lie algebra $\mathfrak{g}$. The space dual to $\mathfrak{g}$ is denoted by $\mathfrak{g^{*}}$. For any $v\in\mathfrak{g^{*}}$ and $X\in\mathfrak{g}$, $\mathfrak{g^{*}}$ is the space of linear maps $v:\mathfrak{g}\rightarrow \mathbb{R}$,
\begin{equation}
\langle v,X\rangle=v(X).
\end{equation}
This pairing may be used to implicitly define the coadjoint action $\mathrm{Ad}^*_g$ of $G$ on $\mathfrak{g^{*}}$,
\begin{equation}
\langle\mathrm{Ad}^{*}_g(v),X\rangle=\langle v, \mathrm{Ad}_{g^{-1}}(X)\rangle.
\label{eq:relation_coadj_adj}
\end{equation}
The space $\mathfrak{g^{*}}$ foliates into symplectic leaves. These symplectic manifolds are the {\it coadjoint orbits}  defined as the set of points related by a coadjoint transformation. Each of these orbits carries a label, which we call $v_0$. We may think of each of these orbits as a phase space. Similar to a classical system, where we define Hamiltonian actions on the phase space, we may define an action that is invariant under $G$ on the coadjoint orbit. This is the { \it  geometric action}. First, let us define the coadjoint orbit $\CO_{v_0}$.  It is given by
\begin{equation}
O_{v_0}=\{v=\mathrm{Ad}^{*}_g(v_0)\,\,|\,\, g\,\in G\}.
\label{eq:orbit_def}
\end{equation}
The manifold is isomorphic to $G/H_{v_0}$, where $H_{v_0}$ is the stabilizer of the orbit given by all elements $h\in H_{v_0}$ that leave $v_0$ invariant under coadjoint transformations.

On these orbits, we may define a symplectic form, the Kirillov-Konstant form, which in turn may be used to define a geometric action on the orbit. This action will depend on a path on the orbit. In the context of complexity, it is more convenient to represent the geometric actions in terms of group elements $g$. Therefore, we use a presymplectic form $\omega$ on $G$ that is related to the Kirillov-Konstant form by a pullback to the orbit. The definition of $\omega$ involves the Maurer-Cartan form $\theta_g$, which maps elements from the tangent space at some point $g$ to those in the tangent space at the identity. In terms of two points $s$ and $t$, it reads
\begin{equation}
\theta_{g}=\left.\frac{d}{ds}\right\rvert_{s=t}\((g^{-1}(t)\cdot(g(s)\).
\label{eq:MC_general_without_central_extension}
\end{equation}
The presymplectic form $\omega$ is then given by
\begin{equation}
\omega=-d\langle v, \theta_g\rangle.
\label{eq:presymplectic_form}
\end{equation}
$\omega$ is closed and locally exact, but in contrast to the Kirillov-Kostant form degenerate. Since it is locally exact, we may define a symplectic potential $\alpha$ by $\omega=d\alpha$. The geometric action on the orbit is then given by
\begin{equation}
I=\int \alpha=-\int_{g(t)} \langle v, \theta_g\rangle = -\int_{g(t)} \langle \mathrm{Ad}^{*}_g(v_0), \theta_g \rangle,
\label{eq:geometric_action_without_central_extension}
\end{equation}
where $g(t)$ is a group path in $G$.

\subsection{Central extension}
\label{sec:central_extensions}
We briefly review how central extensions of Lie groups $G$ arise from projective unitary representations in Hilbert space. For a detailed discussion, we refer to \cite{BMS}.

Let $G$ be a Lie group with Lie algebra $\mathfrak{g}$ and $U_g$ a unitary operator representing a group element $g$ in Hilbert space. In particular, any two operators $U_g$ satisfy $U_{g_1}U_{g_2}=U_{g_1g_2}$. Note that $U_g$ acts on vectors, not on states, the latter being represented by a ray, defined as the set of vectors that only differ by a phase. Therefore, the action of two operators $U_{g_1}$ and $U_{g_2}$ on a vector $\ket{\Phi}$ is ambiguous by a phase,
\begin{equation}
U_{g_1}U_{g_2}\ket{\Phi}=e^{C(g_1, g_2)}U_{g_1 g_2}\ket{\Phi}.
\label{eq:cocycle_rep}
\end{equation}
In case of a non-trivial phase, i.e. a phase that cannot be eliminated by redefining $U_g$, the operators $U_g$ furnish a projective representation of $G$. The central extension of a group $G$ by $S^1$ then arises from this non-trivial phase, and the representation becomes exact for the centrally extended group $\hat{G}=G\otimes\mathbb{R}$. 

Let $V$ be a vector space; in the present context, this is the Hilbert space. Then, the central extension is defined by the phase $C$, which is the 2-cocycle of the group $G$, defined as a map 
\begin{equation}
C:G\times G\rightarrow V:g_1, g_2\rightarrow C(g_1, g_2).
\label{eq:2-cocycle}
\end{equation}
The 2-cocycle satisfies
\begin{equation}
C(g, g^{-1})=0.
\label{eq:cocycle_inverse}
\end{equation}
Group elements of the centrally extended group $\hat{G}$ are pairs $(g, \alpha)$ with $g\in G$ and $\alpha\in\mathbb{R}$. The inverse element is given by
\begin{equation}
(g, \alpha)^{-1}=(g^{-1}, -\alpha).
\end{equation}
The group multiplication is defined by
\begin{equation}
  (g_1,\alpha)(g_2,\beta)=(g_1g_2, \alpha+\beta+C(g_1,g_2)).
  \label{eq:definition_cocycle}
\end{equation}
Thus, for centrally extended groups the product of two group elements always carries an additional phase factor $C(g_1,g_2)$.

\subsection{Geometric action for centrally extended groups}
\label{sec:geo_action_with_central_extension}

The Virasoro and Kac-Moody groups, which we considered in the context of complexity in sec.~\ref{sec:complexity_symmetry_group}, are centrally extended. Moreover, as mentioned in the introduction to sec.~\ref{sec:geometric actions}, the complexity obtained form the complexity measure \eqref{eq:complexity_general_CM} is very similar to the geometric action. In order to better understand this relation, we  need to find a general expression for geometric actions on coadjoint orbits for groups $\hat{G}$ centrally extended by the circle.  
First, we summarize the results of \cite{Alekseev,  Alekseev_3, Alekseev_4, BMS}. Geometric actions for the Virasoro and Kac-Moody groups were considered in \cite{Alekseev,  Alekseev_3, Alekseev_4}. For a detailed discussion of geometric actions in general, we refer to \cite{BMS}. 

From \eqref{eq:orbit_def} and the preceding discussion of central extensions, it is evident that in order to find the centrally extended version of the geometric action \eqref{eq:geometric_action_without_central_extension}, we need to find the coadjoint action of $\hat{G}$ and the central extension of the Maurer-Cartan form. 

Let $(X,m)\in \hat{\mathfrak{g}}$. Elements of the dual Lie algebra $\hat{\mathfrak{g}}^{*}$ are denoted by $(v,c)$. The definition \eqref{eq:relation_coadj_adj} of the coadjoint representation still holds as the central element is invariant under coadjoint transformations.
Moreover, the definition of a coadjoint orbit $\CO_{v_0}$ for centrally extended groups is very similar to \eqref{eq:orbit_def} since we simply replace the coadjoint action of $g\in G$ on $v_0\in\mathfrak{g^{*}}$ with the coadjoint action of $(g,\alpha)\in \hat{G}$ on $(v_0,c)\in\mathfrak{\hat{g^{*}}}$, 
\begin{equation}
O_{v_0}=\{v=\mathrm{Ad}^{*}_{(g, \alpha)} (v_0, c)\,\,|\,\, (g, \alpha)\,\in \hat{G}\}.
\label{eq:orbit_def_new}
\end{equation}
First, let us comment on the coadjoint action $\mathrm{Ad}^{*}_{(g, \alpha)}$ on $v_0$. The central element $\alpha$ acts trivially and thus $\mathrm{Ad}^{*}_{(g, \alpha)}$ reduces to $\mathrm{Ad}^{*}_{g}$. For centrally extended groups $\hat{G}$, $\mathrm{Ad}^{*}_{g}v_0$ consists of two terms, one arising from the the action of $g\in G$ on $v_0$, which we denote by $\mathrm{\widetilde{Ad}}^{*}_g$ and coincides with the definition of \eqref{eq:orbit_def}, the other is a consequence of the central extension and is given by the product of the $1$-cocycle $\gamma(g)$ of the group $G$ and the central element $c$,
\begin{equation}
\mathrm{Ad}^{*}_{g}(v_0,c)=(\mathrm{\widetilde{Ad}}^{*}_g(v_0)+c\gamma(g),c).
\label{eq:coadjoint_general}
\end{equation}

Next, we define the centrally extended Maurer-Cartan form. It is given by
\begin{equation}
(\theta_{(g, \alpha)},m_\theta)=\left.\frac{d}{ds}\right\rvert_{s=t}\((g^{-1}(t),-\beta(t))\cdot(g(s),\, \beta(s)) \),
\label{eq:MC_general}
\end{equation}
where $f(t)$ and $\beta(t)$ are paths in the group manifold of $G$ and the real numbers, respectively. Note that the only difference to \eqref{eq:MC_general_without_central_extension} is the additional path $\beta(t)$.
Using \eqref{eq:definition_cocycle}, \eqref{eq:MC_general} may be written as
\begin{equation}
(\theta_{(g, \alpha)},m_\theta)=\left.\frac{d}{ds}\right\rvert_{s=t}\(g^{-1}(t)\circ g(s), -\beta(t)+\beta(s)+C(g^{-1}(t), g(s))\).
\label{eq:MC_form}
\end{equation}

Then, the geometric action \eqref{eq:geometric_action_without_central_extension} on an orbit specified by $v_0$ generalized to groups centrally extended by $S^1$ reads with \eqref{eq:orbit_def_new}, \eqref{eq:coadjoint_general}, and \eqref{eq:MC_form}
\begin{equation}
I=-\int_g\langle v, (\theta,m_\theta)\rangle=-\int_g \,\(\langle\mathrm{\widetilde{Ad}^{*}}(v_0),\theta_g\rangle+c\langle\gamma(g),\theta_g\rangle +cm_\theta \). 
\label{eq:general_coadjoint_action}
\end{equation} 
Compared to the geometric action \eqref{eq:geometric_action_without_central_extension} for non-centrally extended groups, here we have additional terms $c\langle\gamma(g),\theta_g\rangle$ and $cm_\theta$.
We will show in \secref{sec:complexity and geometric actions} that the difference between the geometric action and the complexity functional considered in \secref{sec:complexity_general} is due to these terms.

\subsection{Geometric actions for Virasoro and Kac-Moody groups}
\label{sec:geo_Vir_Kac}
We now apply the result for the geometric action of centrally extended groups given by \eqref{eq:general_coadjoint_action} to the Virasoro and Kac-Moody groups based on \cite{Alekseev,  Alekseev_3, Alekseev_4}. Furthermore, the geometric action of the Virasoro group has been considered in the context of Virasoro Berry phases in \cite{Berry_phases}, which also includes a review of Virasoro coadjoint orbits and its cocycle. For a review of Kac-Moody geometric actions, see also e.g \cite{Barnich}, from which we adopt the notation. 

The results of this section will enable us to spot the similarities between geometric actions for these groups and the complexity functionals \eqref{eq:complexity_Vir} and \eqref{eq:complexity_Kac-Moody_CM} in sec.~\ref{sec:complexity and geometric actions}.

\subsubsection{Geometric action for the Virasoro group}
\label{sec:geometric_action_Vir}
Here, we apply the result \eqref{eq:general_coadjoint_action} for the geometric action in the presence of a central term as presented in sec.~\ref{sec:geo_action_with_central_extension} to the Virasoro group.\\
From sec.~\ref{sec:central_extensions}, we know that the central term $C(g_1,g_2)$ in \eqref{eq:definition_cocycle} is nothing but the 2-cocycle defining the central extension. It was first derived in \cite{Bott} for the inverse diffeomorphism $F$ and reads
\begin{equation}
C(F_1, F_2)=-\frac{1}{48\pi}\int_0^{2\pi}d\sigma\, \mathrm{log}\(F^{\prime}_1(F_2(\sigma))\)\frac{F_2^{\prime\prime}(\sigma)}{F_2^{\prime}(\sigma)}
\label{eq:cocycle_Vir}
\end{equation}
for $F_1, F_2\in \mathrm{Diff}^+( S^1)$. 
To derive the geometric action \eqref{eq:general_coadjoint_action}, we need the coadjoint action of group elements $F$ and the Maurer-Cartan form. Note that we use the inverse diffeomorphism $F$ since the similarities with the complexity functional \eqref{eq:complexity_Vir} are easier to see this way. 
Let $(b, c)=(b(\sigma)d\sigma^2, c)\in \mathfrak{\hat{g}^*}$, where $\mathfrak{\hat{g}^*}$ denotes the dual space of the Virasoro algebra.
Then, the coadjoint action of $F$ on $(b_0, c)$ is given by
\begin{equation}
\mathrm{Ad^{*}}_F(b_0,c)=\(b_0(F(\sigma))(\partial_\sigma F(\sigma))^2-\frac{c}{24\pi}\{F, \sigma\},c\).
\label{eq:Vir_coadjoint}
\end{equation}
By comparison of \eqref{eq:Vir_coadjoint} with \eqref{eq:coadjoint_general}, the 1-cocycle of the Virasoro group may be identified as $\gamma(F)=-\frac{1}{24\pi}\{F, \sigma\}$.

Moreover, the Maurer-Cartan form of the Virasoro group may be derived from \eqref{eq:MC_form} and reads
\begin{equation}
(\theta, m_\theta)=\(\frac{\dot{F}dt}{F^{\prime}}\frac{\partial}{\partial \sigma},\frac{1}{48\pi}\int_0^{2\pi}d\sigma \frac{\dot{F}dt}{F^{\prime}}\(\frac{F^{\prime\prime}}{F^{\prime}}\)^{\prime}\). 
\label{eq:Vir_Maurer-Cartan}
\end{equation}
In particular, the central term $m_\theta$ is obtained from the cocycle \eqref{eq:cocycle_Vir} according to \eqref{eq:MC_form}.
To obtain the geometric action for the Virasoro group, we may then simply insert \eqref{eq:Vir_coadjoint} and \eqref{eq:Vir_Maurer-Cartan} into \eqref{eq:general_coadjoint_action} and obtain
\begin{equation}
I=\int d\sigma dt\(-b_0(F)F^\prime\dot{F}+\frac{c}{48\pi}\frac{\dot{F}}{F^{\prime}}\(\frac{ F^{\prime\prime\prime}}{F^{\prime}}-2\frac{(F^{\prime\prime})^2}{(F^\prime)^2}\) \).
\label{eq:geometric_action_Vir}
\end{equation}
This result is essential for showing the connection between geometric actions and complexity functionals, as will be done  in sec.~\ref{sec:complexity and geometric actions}.
\\
\subsubsection{Geometric action for the Kac-Moody group}
\label{sec:Geometric_Action_Kac-Moody}
We now turn to our second example of Kac-Moody groups and apply the result \eqref{eq:general_coadjoint_action} of sec.~\ref{sec:geo_action_with_central_extension} above. According to sec.~\ref{sec:central_extensions}, the central term $C(g_1,g_2)$ in \eqref{eq:definition_cocycle} is given by the 2-cocycle defining the central extension of the loop group and thus the Kac-Moody group. It was first derived in \cite{Mickelsson} and reads
\begin{equation}
C(g_1, g_2)=\frac{1}{4\pi}\int_{D}\,\tr\left[ g_1^{-1}\bar{d}g_1\bar{d}g_2g_2^{-1}\right].
\label{eq:cocycle_Kac_Moody}
\end{equation}
Note that the last integral is over the disk $D$ with boundary $S^1$, parametrized by $\sigma$, and $r\in [0,1]$. $\bar{d}$ denotes the the exterior derivative with respect to these coordinates.
Next, we define the coadjoint action. Consider a pair $(v, k)\in\mathfrak{\hat{g}^*}$. Then, the coadjoint action of a group element $g$ on $(v_0, k)$ is given by
\begin{equation}
\mathrm{Ad}^{*}_g(v_0, k)=\(gv_0g^{-1}-\frac{k}{2\pi} \partial_{\sigma}gg^{-1}, k\).
\label{eq:coadjoint_Kac-Moody}
\end{equation}
By comparison of \eqref{eq:coadjoint_Kac-Moody} with \eqref{eq:coadjoint_general}, it becomes evident that the 1-cocylce of the Kac-Moody group is given by $\gamma(g)=-\frac{1}{2\pi} \partial_{\sigma}gg^{-1}$.

Similar to sec.~\ref{sec:geometric_action_Vir}, the Maurer-Cartan form for the Kac-Moody group may be derived from \eqref{eq:MC_form}. Since according to \eqref{eq:MC_form}, the central extension of the Maurer-Cartan form is defined in terms of the cocycle \eqref{eq:cocycle_Kac_Moody}, it inherits the special feature  \eqref{eq:cocycle_Kac_Moody} that its central term lives on a disk with boundary $S^1$ rather than $S^1$ itself. As mentioned in sec.~\ref{sec:geometric_action_Vir} for the Virasoro group, the relation between the central extension of the Maurer-Cartan form and the cocycle implies for the Kac-Moody group as well that the central extension of the Maurer-Cartan form will be essential in obtaining computational cost for the central term $C(g_1,g_2)$ in \eqref{eq:definition_cocycle}.

Let $M$ denote the manifold given by $\gamma\times D$, where $D$ is the disk with boundary $S^1$ parametrized by $\sigma$ and $r\in [0,1]$ as above, and $\gamma$ is a curve parametrized by $t$. With the notation $d=dt\partial_t$, the left-invariant Maurer-Cartan form reads
\begin{equation}
(\theta_g, m_\theta)=\(g^{-1}dg, \frac{1}{4\pi}\bigg[\int_{0}^{2\pi} d\sigma\,\tr\left[ g^{-1}\partial_{\sigma}gg^{-1}dg\right]+\int_{D} \tr\left[ g^{-1}(\bar{d}g g^{-1})^2d g \right]\bigg]\).
\label{eq:MC_Kac-Moody_literature}
\end{equation}
The geometric action is then simply obtained by inserting \eqref{eq:coadjoint_Kac-Moody} and \eqref{eq:MC_Kac-Moody_literature} into \eqref{eq:general_coadjoint_action},
\begin{equation}
I=\int d\sigma\, \tr\bigg[-v_0dgg^{-1}+\frac{k}{4\pi}dg g^{-1}\partial_{\sigma} g g^{-1}\bigg]-\frac{k}{4\pi}\int_M\tr\bigg[d g g^{-1} \(\bar{d}g g^{-1}\)^2\bigg].
\label{eq:geometric_action}
\end{equation}
In the first two terms $g\equiv g(\sigma, r=1, t)$, implying the integral is over the boundary of the manifold $M$, $\partial M= \gamma \times S^1$, whereas in the last term we integrate over $M$. 
Moreover, the last term is the topological Wess-Zumino term. Note, in particular, that for $v_0=0$, \eqref{eq:geometric_action} is just the chiral WZW action. This was observed in \cite{Alekseev} in the context of geometric quantization. 

Both \eqref{eq:geometric_action_Vir} and \eqref{eq:geometric_action} play a central role when considering complexity for these groups as we explain below.

\section{From complexity to geometric actions}
\label{sec:complexity and geometric actions}
Based on the results presented in secs.~\ref{sec:complexity_general} and \ref{sec:geometric actions}, here we propose a modified cost function which yields the geometric action as a complexity functional.
We begin in \secref{sec:relation} by comparing contributions to the complexity functionals \eqref{eq:complexity_Vir} and \eqref{eq:complexity_Kac-Moody_CM} to those occurring in the geometric actions \eqref{eq:geometric_action_Vir} and \eqref{eq:geometric_action}.
We find that we may identify the velocities \eqref{eq:velocity_Vir} and \eqref{eq:velocity_Kac-Moody} with the non-centrally extended Maurer-Cartan form \eqref{eq:MC_general_without_central_extension}.
The transformations \eqref{eq:transformation_T} and \eqref{eq:transformation_current} acting on the symmetry currents $T(\sigma)$, $J(\sigma)$ are identified with the coadjoint action of the according group.
Then it becomes evident that the complexity described in sec.~\ref{sec:complexity_general} is not sensitive to contributions from the central extension, since there is manifestly no contribution of the central extension to \eqref{eq:complexity_general_CM}.

To obtain the full geometric action including the central extension term as our complexity functional we introduce a generalized version of the gate equation \eqref{eq:velocity} in sec. \ref{sec:Recovering_the_central_term}.
From this new gate equation, the centrally extended Maurer-Cartan form \eqref{eq:MC_general} is derived, independently of the specific group.
This allows us to derive in sec.~\ref{sec:geometric_Action=complexity} that the new complexity expression obtained in this way is equal to the geometric action of the corresponding group.

\subsection{Relation between complexity and geometric action}
\label{sec:relation}
Here we show that the geometric action is equal to the complexity \eqref{eq:complexity_general_CM} up to terms arising from the central extension of the Maurer-Cartan form by identifying corresponding elements in the geometric actions \eqref{eq:geometric_action_Vir} and \eqref{eq:geometric_action} with those in the complexities \eqref{eq:complexity_Vir} and \eqref{eq:complexity_Kac-Moody_CM}.

The results \eqref{eq:complexity_Vir} and \eqref{eq:complexity_Kac-Moody_CM} for the complexity may be summarized as follows. Given a highest-weight reference state $\ket{\psi}$, a unitary representation $U_g$ of a symmetry group $G$ and an operator $J$, corresponding to the conserved currents of the CFT under the symmetry transformations $U_g$, the complexity functional reads
\begin{equation}
\CC=\frac{1}{2\pi}\int_{0}^{T}dt\int_{0}^{2\pi}\epsilon(t, \sigma)\bra{\psi}U^{\dagger}_gJ U_g\ket{\psi}.
\label{eq:complexity_general}
\end{equation}
For the Virasoro group, we identify $g=f$, for the Kac-Moody group $g=\Omega$.

We now  show that \eqref{eq:complexity_general} is equal to the geometric action without contributions from the central extension of the Maurer-Cartan form,
\begin{equation}
\CC \stackrel{!}{=} -\int \langle \mathrm{Ad}^{*}_{g}(v_0), \theta_{g}\rangle.
\label{eq:complexity_geometric_action_missing_central_term}
\end{equation}
If this holds, we may identify the transformed current with the coadjoint action of group elements on $v_0\in \mathfrak{g}^*$, where the orbit is specified by the expectation value of the conserved currents $J$ of the according group. Furthermore, the Mauer-Cartan form $\theta_{g^{-1}}$ in terms of the inverse group element corresponds to the velocities $\epsilon$. 

The equality between \eqref{eq:complexity_general} and \eqref{eq:complexity_geometric_action_missing_central_term} may be obtained directly from the gate equation \eqref{eq:velocity}.
This is possible by rewriting \eqref{eq:velocity} in terms of two different points $t$ and $s$ in time that do not necessarily have to be close. Then, the symmetry transformation $g(t)$ is given by
\begin{equation}
g(t)=e^{\int_{s}^{t}\epsilon(s^\prime)ds^\prime}g(s).
\label{eq:generalized_velocity}
\end{equation}
Upon multiplication by the inverse $g^{-1}(s)$ and performing a derivative with respect to $s$, we obtain
\begin{equation}
\left.\frac{d}{ds}\right\rvert_{s=t}(g(t) g^{-1}(s))=-\epsilon(t).
\end{equation}
This is the expression for the non-centrally extended Maurer-Cartan form $\theta_{g^{-1}}$ given in \eqref{eq:MC_general_without_central_extension} in terms of $g^{-1}$ rather than $g$.
Hence it becomes clear that \eqref{eq:complexity_general} and \eqref{eq:complexity_geometric_action_missing_central_term} are in fact equal to each other.
To see this more explicitly, we now proceed by identifying corresponding contributions in both actions for the examples of the Kac-Moody and Virasoro groups, making use of the results of the preceding sections.

\subsubsection*{Virasoro group}
For the Virasoro group, the transformation of the energy-momentum tensor under symmetry transformations $U_f$, expressed in terms of the inverse diffeomorphism $F$ as given in \eqref{eq:transformation_T}, is directly equivalent to the coadjoint action \eqref{eq:Vir_coadjoint} up to an overall sign and a prefactor if we identify $v_0 = b_0(F)=-\frac{1}{2\pi}\bra{h}T(F)\ket{h}$,
\begin{equation}
\frac{1}{2\pi}\bra{h}(U^{\dagger}_{f}T(\sigma)U_f)(F)\ket{h}=-\mathrm{Ad}^{*}_F(b_0).
\end{equation}
Moreover, the component of the Maurer-Cartan form \eqref{eq:Vir_Maurer-Cartan} without central element is equal up to a sign to the velocities $\epsilon$ computed in \eqref{eq:velocity_Vir},
\begin{equation}
\epsilon(t, \sigma)=-\theta_F.
\label{eq:identification_Vir_velocity}
\end{equation}

\subsubsection*{Kac-Moody group}
Similarly, by comparison of \eqref{eq:coadjoint_Kac-Moody} with \eqref{eq:transformation_current}, we find that upon identifying $g\rightarrow\Omega$ and $v_0 = -\frac{1}{2\pi}\bra{\Phi_{\lambda}}J\ket{\Phi_{\lambda}} = -\frac{1}{2\pi}C_{\lambda}$, the coadjoint action is just the transformation property of the currents $J$ under symmetry transformations,
\begin{equation}
\frac{1}{2\pi}\bra{\Phi}U^{\dagger}J(\sigma)U\ket{\Phi}=-\mathrm{Ad}^{*}_g(v_0).
\label{eq:identification_exp_J_v_0}
\end{equation} 
By comparing the velocity \eqref{eq:velocity_Kac-Moody} with the Maurer-Cartan form \eqref{eq:MC_Kac-Moody_literature}, we identify the Maurer-Cartan form of the inverse element $g^{-1}$ with the velocity. Note that we have already done the identification in terms of the inverse group element automatically in the Virasoro case as the velocities are functions of the inverse diffeomorphism $F$. For the Kac-Moody group, the replacement $g \to g^{-1}$ leads to an overall sign in $\theta_g$ in \eqref{eq:MC_Kac-Moody_literature} since $\theta_{g^{-1}}=g\partial_{t}g^{-1}dt=-\partial_tgg^{-1}dt$. Therefore, just as in the Virasoro case, the component of the Maurer-Cartan form $\theta_{g^{-1}}$ without central term is equal up to a sign to the velocities computed in \eqref{eq:velocity_Kac-Moody},
\begin{equation}
\epsilon(t,\sigma)=-\theta_{g^{-1}}.
\end{equation}
Thus for both the Virasoro and Kac-Moody group, we see directly that the complexity functional \eqref{eq:complexity_general} is equal to the geometric action \eqref{eq:complexity_geometric_action_missing_central_term} without the central extension term.

Let us comment on why it is necessary to write the Maurer-Cartan form in terms of the inverse group element $g^{-1}$ (and similarly in terms of the inverse diffeomorphism $F$ in \eqref{eq:identification_Vir_velocity}). This is easily understood from the complexity point of view. By definition, the Maurer-Cartan form relates a velocity at some point $g(t)$ to that at the identity. However, when computing the complexity, we start at the identity and aim at relating the velocity at $t=0$ to the velocity at some point close by. We are thus moving "forward" along the path, whereas the Maurer-Cartan form by definition moves us "backward". Consequently, the Maurer-Cartan form in the complexity calculation must be written in terms of the inverse transformation in order to move "forward". 

\subsection{Recovering the central term}
\label{sec:Recovering_the_central_term}
The question now arises if it is possible to adjust the cost function \eqref{eq:1_norm_cost_function} to obtain the full geometric actions \eqref{eq:geometric_action_Vir} and \eqref{eq:geometric_action} for the complexity functionals \eqref{eq:complexity_Vir} and \eqref{eq:complexity_Kac-Moody_CM}. To this end, we have to find an expression for the velocities that is sensitive to contributions from the central extension. This is accomplished by adding a new contribution to that of the path through the symmetry transformations. The additional contribution is a path in the real numbers that is entirely defined in terms of the chosen path through the symmetry transformations.

The starting point is \eqref{eq:generalized_velocity}, which, as a reminder, is the generalized version of $\eqref{eq:velocity}$ for two points $t$ and $s$ along the path that do not necessarily have to be close.
Rather than just considering a path through the transformations $g$, we also allow a path through the real numbers $\alpha(t)$,
\begin{equation}
(g(t), \alpha(t))=e^{\int_s^tds^\prime(\epsilon(s^\prime), \beta(s^\prime))}(g(s), \alpha(s)).
\label{eq:new_path}
\end{equation}
To derive the corresponding velocities, we multiply by $(g(s), \alpha(s))^{-1}$ and take a derivative with respect to $s$. This yields the central extension of the Maurer-Cartan form according to \eqref{eq:MC_general},
\begin{equation}
(\theta, m_\theta)_{g^{-1}}=-(\epsilon(t), \beta (t))=\left.\frac{d}{ds}\right\rvert_{s=t}\((g(t),\alpha(t))\cdot(g^{-1}(s),\, -\alpha(s)) \).
\label{eq:MC_not_explicit}
\end{equation}
With \eqref{eq:definition_cocycle}, \eqref{eq:MC_not_explicit} then becomes
\begin{equation}
(\theta, m_\theta)_{ g^{-1}}=-(\epsilon(t), \beta (t))=\(\left.\frac{d}{ds}\right\rvert_{s=t}g(t)\circ g^{-1}(s), \left.\frac{d}{ds}\right\rvert_{s=t}C(g(t)g^{-1}(s))\),
\label{eq:MC_central_extension}
\end{equation}
Then, the new cost function that we define for the velocities $\epsilon(t)$ and $\beta(t)$ in the extended group is given by a sum of two contributions,
\begin{equation}
  \CF=\int d\sigma \epsilon(t,\sigma)|\bra{\psi_R }U^{\dagger}(t)J U(t)\ket{\psi_R}| + c_0\beta(t),
  \label{eq:new_cost_function}
\end{equation}
where $J$ is the conserved current of the symmetry group in consideration.
The first contribution from $\epsilon(t)$ is just the cost function \eqref{eq:1_norm_cost_function} used previously, which does not include the central extension.
The second contribution is given by $\beta(t)$ times a constant $c_0$.
We identify $c_0$ with the Virasoro central charge $c$ or the level $k$ of the Kac-Moody group, respectively,  to achieve an equality of the resulting complexity functional with the geometric actions for the corresponding symmetry groups (see \eqref{eq:complexity=geometric action Vir} and \eqref{eq:complexity=geometric action Kac-Moody} below).

Two comments are in order.
First, we note that we have chosen the real number valued part $\alpha$ of the group path to be a constant, $\alpha(t)=\alpha(s)=\mathrm{const}$.
This choice is necessary to avoid those contributions that are solely determined by the path through the real numbers and are thus independent of the transformation $g$ applied.
The cost, however, should always depend on the symmetry transformation.
Hence, the only contribution from the central extension should arise from the cocycle $C(g_1,g_2)$, which depends on the transformations $g$ alone, and not from derivatives of $\alpha$, which are independent of $g$.
Moreover, derivatives of $\alpha$ would lead to additional contributions that do not occur in geometric actions.
Since we aim to show that we can find a cost function such that the complexity equals the geometric action, we need to choose $\alpha = \mathrm{const}$.  

Second, before we proceed we have to verify that the cost function \eqref{eq:new_cost_function} still assigns zero cost to the identity. This is easily checked by considering \eqref{eq:new_path} for identical points along the path such that
\begin{equation}
(g(s), \alpha(s))=e^{\int_{s}^{s}ds^\prime(\epsilon(s^\prime), \beta(s^\prime))}(g(s), \alpha(s)),
\end{equation}
which implies that $\int_{s}^{s}ds^\prime(\epsilon(s^\prime), \beta(s^\prime))$ has to vanish. This is obviously satisfied since the integration interval is of zero size. From a physical point of view, we may also argue that for two identical points $t=s$, we expect the velocities \eqref{eq:MC_central_extension} to vanish. This is satisfied as a consequence of the properties of the Maurer-Cartan form \eqref{eq:MC_central_extension}: For the non-centrally extendend component $\epsilon(t)$, we take the derivative of a constant since $g(s)g^{-1}(s)=1$. Thus, $\epsilon(s)=\theta_{g^{-1}}=0$. Similarly, for the central extension piece $\beta(t)=m_{\theta}=0$ if $C(g(s)g^{-1}(s))=\mathrm{const}$. Since according to \eqref{eq:cocycle_inverse}, the cocycle vanishes for the argument $g(s)g^{-1}(s)$, the new cost function does indeed assign zero cost to the identity transformation. This further justifies setting $\alpha=\mathrm{const}$, as otherwise we would have obtained an additional derivative.

\subsection{Complexity=geometric action}
\label{sec:geometric_Action=complexity}
We now use the definitions of \secref{sec:Recovering_the_central_term} to determine the complexity resulting from the new gate equation \eqref{eq:new_path}.
We express the obtained complexity functionals entirely in terms of the coadjoint action and the Maurer-Cartan form.

First, we compute the centrally-extended velocities and thus the Maurer-Cartan form in terms of the inverse $g^{-1}$ as given in \eqref{eq:MC_central_extension}.
For the Virasoro group, the component without central extension can be directly obtained from
\begin{equation}
\theta=\left.\frac{d}{ds}\right\rvert_{s=t}f(t)\circ F(s).
\end{equation}
For the central extension, we use the 2-cocycle of the Virasoro group \eqref{eq:cocycle_Vir},
\begin{equation}
m_\theta=\left.\frac{d}{ds}\right\rvert_{s=t}C(f(t),F(s))=-\frac{1}{48\pi}\int_{0}^{2\pi}\left.\frac{d}{ds}\right\rvert_{s=t}\(\mathrm{log}\left[f^{\prime}(t,(F(s)))\right]\frac{F^{\prime\prime}(t)}{F^{\prime}(t)}\).
\end{equation}
The result reads (see e.g. \cite{Berry_phases})
\begin{equation}
(\theta, m_\theta)=\(\frac{\dot{F}}{F^{\prime}}, \frac{1}{48\pi}\int_{0}^{2\pi}d\sigma\frac{\dot{F}}{F^{\prime}}\( \frac{F^{\prime\prime}}{F^{\prime}}\)^{\prime}\).
\label{eq:MC_form_Vir_gate}
\end{equation}
As expected, this coincides with \eqref{eq:Vir_Maurer-Cartan}\footnote{Strictly speaking, they differ by $dt$ since from the gate equation \eqref{eq:new_path}, we just obtain the components of the Maurer-Cartan form. The missing differential $dt$ is absorbed in the integral of the computational cost (see e.g. \eqref{eq:complexity_general_CM}).}, verifying that the modified gate equation \eqref{eq:new_path} yields the centrally extended Maurer-Cartan form.

For the Kac-Moody group, the Maurer-Cartan form without central extension reads
\begin{equation}
\theta_{g^{-1}}=\left.\frac{d}{ds}\right\rvert_{s=t}\,g(t)g^{-1}(s)=-\partial_t g(t)g^{-1}(t).
\label{eq:MC_form_KM_gate}
\end{equation}
The central extension can be computed from the 2-cocycle of the loop group \eqref{eq:cocycle_Kac_Moody},
\begin{equation}
m_\theta=\left.\frac{d}{ds}\right\rvert_{s=t}C(g(t),g^{-1}(s))=\frac{1}{4\pi}\int \tr\left[g^{-1}(t)\bar{d}g(t)\left.\frac{d}{ds}\right\rvert_{s=t}(\bar{d}g^{-1}(s)g(s))\right].
\end{equation}
This gives
\begin{equation}
m_\theta=-\frac{1}{4\pi}\int d\sigma\, \tr\left[\partial_{\sigma }gg^{-1}\partial_t gg^{-1}\right]-\frac{1}{4\pi}\int \tr\left[(\bar{d}gg^{-1})^2\partial_t gg^{-1}\right].
\label{eq:MC_form_KM_gate_central_term}
\end{equation}
By comparison with \eqref{eq:MC_Kac-Moody_literature}, we find that $(\theta_{g^{-1}}, m_{\theta_{g^{-1}}})=-(\theta_g, m_{\theta_g})$.

\medskip
Based on this result, the centrally-extended Maurer-Cartan forms \eqref{eq:MC_form_Vir_gate} and \eqref{eq:MC_form_KM_gate_central_term} arising from the new gate equation \eqref{eq:new_path} now enable us to obtain the full geometric action.
Let $\ket{\psi}$ be a highest-weight state. Then, with the coadjoint action  \eqref{eq:Vir_coadjoint} and the Maurer-Cartan form \eqref{eq:MC_form_Vir_gate}, the complexity for the Virasoro group with contributions from the central extension may be written as
\begin{equation}
  \CC=\int dt\int d\sigma\, \(\epsilon(t), \beta(t)\)\(\frac{1}{2\pi}\bra{\psi}U^{\dagger}_fT U_f\ket{\psi},c\)=-\int dt \(\langle\mathrm{Ad}^{*}_F, \theta_F\rangle+cm_{\theta_{F}}\)=I[F],
  \label{eq:complexity=geometric action Vir}
\end{equation}
where $I[F]$ is the geometric action \eqref{eq:geometric_action_Vir} for the Virasoro group.
For the Kac-Moody group, it is not necessary to express the action in terms of the inverse path $g^{-1}$ due to $(\theta_{g^{-1}}, m_{\theta_{g^{-1}}})=-(\theta_g, m_{\theta_g})$. With the coadjoint action \eqref{eq:coadjoint_Kac-Moody} and the Maurer-Cartan form \eqref{eq:MC_Kac-Moody_literature}, we directly  find
\begin{equation}
  \CC=\int dt\int d\sigma\, (\epsilon(t), \beta(t))\(\frac{1}{2\pi}\bra{\psi}U^{\dagger}_gJ U_g\ket{\psi},k\)=-\int dt \(\langle\mathrm{Ad}^{*}_g(b_0), \theta_g\rangle+km_{\theta_{g}}\)=I[g].
  \label{eq:complexity=geometric action Kac-Moody}
\end{equation} 
Here, $I[g]$ is the geometric action \eqref{eq:geometric_action} for the Kac-Moody group.
In terms of group theory language, the relations \eqref{eq:complexity=geometric action Vir} and \eqref{eq:complexity=geometric action Kac-Moody} may be summarized as follows. The complexity for a CFT with symmetry group $G$ and group elements denoted by $g$ is given by
\begin{equation}
\CC=I[g]=-\int dt\,\(\langle\mathrm{Ad}^{*}_g(b_0), \theta_g\rangle+km_{\theta_{g}}\),
  \label{eq:complexity=geometric action general}
\end{equation}
where $\mathrm{Ad}^{*}_g(b_0)$ contains the information about the transformation of the conserved currents under the according symmetry transformations. In particular, $b_0$ labels the coadjoint orbit and may be identified with the expectation value of the conserved current. The Maurer-Cartan form $(\theta_{g}, m_{\theta_{g}})$ corresponds to the velocity in terms of the path $g$.
In summary, the map between group theory quantities and complexity quantities that we established implies the following: For the modified cost function \eqref{eq:new_cost_function}, the complexity for a particular centrally extended group coincides exactly with the on-shell value of the geometric action of this group.

For the Virasoro group, the similarities between the coadjoint action and the complexity are most obvious when the latter is written as a function of the inverse $F$. However, the path \eqref{eq:velocity} considered to find the complexity is a diffeomorphism $f(t, \sigma)$. Therefore, in order to perform explicit calculations of the complexity, it is more convenient to rewrite the geometric action and hence the complexity in terms of $f$. This can be easily accomplished by setting $\sigma=f(t, \tilde{\sigma})$, where $\tilde{\sigma}=F(t, \sigma)$. The result reads
\begin{equation}
\CC=\int dt d\sigma\, \frac{\dot{f}}{f^{\prime}}\(b_0(f)+\frac{c}{48\pi}\(\frac{f^{\prime\prime}}{f^{\prime}}\)^{\prime}\).
\label{eq:complexity_Virasoro_with_central_term}
\end{equation}
Note that this is exactly the same result already anticipated in \cite{Caputa:2018kdj}.
For Kac-Moody groups, we simply set $g\rightarrow\Omega$, and thus,
\begin{equation}
\CC=\int dt d\sigma\, \tr\bigg[-v_0\dot{\Omega}\Omega^{-1}+\frac{k}{4\pi}\dot{\Omega} \Omega^{-1} \Omega^{\prime} \Omega^{-1}\bigg]-\frac{k}{4\pi}\int_M dt\,\tr\bigg[\dot{\Omega} \Omega^{-1} \(\bar{d}\Omega \Omega^{-1}\)^2\bigg].
\label{eq:complexity_Kac-Moody}
\end{equation}
Therefore, we have explicitely shown that the complexity functionals defined in this section are equivalent to the geometric actions derived in \secref{sec:geometric actions} for both the Virasoro and Kac-Moody groups.

\subsection{Gauge invariance}
\label{sec:Gauge invariance}
There is an important subtlety to the interpretation of geometric actions as complexity functionals that we have not yet addressed.
The issue is that not all symmetry transformations lead to physically distinguishable states.
Some transformations, for example $\ket h \to e^{ia L_0}\ket h$, lead only to a phase change.
Thus, applying a certain symmetry transformation $f$ should be equivalent to applying $f$ together with such a phase changing (gauge) transformation.
As it turns out, the on-shell value of the geometric action is not invariant under such gauge transformations.

In the language of coadjoint orbits, the gauge transformations are represented by transformations which leave the orbit invariant.
The coadjoint orbits are isomorphic to manifolds $\hat{G}/H_{v_0}$, where $H_{v_0}$ is the set of transformations that leave the reference state invariant called the \textit{stabilizer} of the orbit \cite{BMS}.
In terms of the the coadjoint element $v_0$ defining the reference state,
\begin{equation}
  \mathrm{Ad}^*_h v_0=v_0 \text{~for~} h\in H_{v_0}.
\end{equation}
The coadjoint orbit action and with it the complexity should be invariant under transformations $h(t)$ from the stabilizer subgroup.
As discussed in sec.~\ref{sec:geometric_action}, the geometric actions are defined not on the full group manifold $\hat G$, but only on the orbit $\hat G/H_{v_0}$.
In particular, only the projection of the group path $g(t)$ onto the coadjoint orbit is relevant for the complexity.
Consequently, paths that are non-trivial in $\hat{G}$ should result in vanishing complexity if they are trivial projections on the coadjoint orbit.
This includes paths $h(t)$ belonging to the stabilizer since these are projected onto points and thus always have vanishing complexity.

It turns out that this gauge invariance property is indeed satisfied -- up to the appearance of a total derivative term in the action \cite{BMS}.
This term does not change the equations of motion.
However, for the on-shell value of the action, this total derivative term leads to an additional contribution, such that the on-shell action is not gauge invariant.
Our complexity definition, however, is defined in terms of the on-shell action. As such,  it is assigned a role as a physically meaningful and in principle measureable object.
Therefore we cannot neglect the boundary terms in the complexity expression.

As an example, we consider the Virasoro group, for which the reference states  fall into two different classes, $\ket{0}$ and $\ket{h}$ with $h > 0$.
While the first is invariant under $SL(2,R)$ transformations generated by $L_{0,\pm 1}$, the latter are invariant only under the $U(1)$ subgroup of the Virasoro group generated by $L_0$.
The corresponding orbits are $\mathrm{Diff}(S^1)/SL(2,R)$ and $\mathrm{Diff}(S^1)/U(1)$, respectively.\footnote{
For a detailed discussion of Virasoro orbits, we refer to \cite{Witten}.}

Now consider a gauge transformation given by an
$U(1)$ stabilizer parametrized by $a(t)$,
\begin{equation}
  f(t,\sigma) \to f(t,\sigma) + a(t).
\end{equation}
Since $a(t)$ is independent of $\sigma$, it leads to an action of the $L_0$ generator only.
Inserting this in the action \eqref{eq:complexity_Virasoro_with_central_term}, we obtain a change of
\begin{equation}
  S \to S + \delta S \text{~with~} \delta S = \int dt d\sigma \frac{\dot a}{f'}\(b_0 + \frac{c}{48\pi}\(\frac{f''}{f'}\)'\).
  \label{eq:gauge transformaed coadjoint action}
\end{equation}
By expressing \eqref{eq:gauge transformaed coadjoint action} in terms of the inverse diffeomorphism $F$, $\delta S$ can be integrated in $\sigma$.
Since $F(\sigma+2\pi,t) = F(\sigma,t) + 2\pi$, we only get a constant contribution from the integral over $\sigma$.
The remaining integral is over a constant times $\dot a$.
Therefore, it is trivially solvable to obtain
\begin{equation}
  S \to S + 2\pi b_0(a(T)-a(0)).
\end{equation}
Thus, it is possible that symmetry transformations which are pure gauge and change only the phase of the reference state lead to a non-vanishing complexity\footnote{Note that this is independent of whether one takes into account the central extension in the cost function. Eq.~\eqref{eq:complexity_CM_Vir} suffers from the same problem.}.
In \secref{sec:Examples}, more examples of transformations with the same property will be encountered.

A further important aspect related to the stabilizer group and boundary terms of the action is the relation of the complexity to Berry phases on the Virasoro coadjoint orbits, which have recently been studied in \cite{Berry_phases}.
In particular, that the complexity is sensitive to phase changes in the reference state may be deduced without any calculation from their relation to Berry phases.
We briefly review the results of \cite{Berry_phases} and highlight the relation to the Virasoro complexities in app. \ref{app:Berry_phases}.

\section{Constructing optimal transformations and computing complexity}
\label{sec:Examples}
In this section, we will provide some intuition for the rather abstract complexity functionals obtained in sec.~\ref{sec:geometric_Action=complexity} by calculating complexity and target states for examples of optimal transformations. This is done for the Virasoro group in sec.~\ref{sec:Ex_Vir} and for the Kac-Moody group in sec.~\ref{sec:Ex_Kac}.

In order to identify optimal transformations for each of these groups, we determine the equations of motion and their solutions for the Virasoro and Kac-Moody complexity functionals \eqref{eq:complexity_Virasoro_with_central_term} and \eqref{eq:complexity_Kac-Moody}, respectively. The solutions then provide the full set of optimal transformations for the respective group. 

Upon calculating the complexity from these solutions, for the Virasoro group we encounter several issues. Not only do we find a large class of optimal transformations for which the Virasoro complexity counts only phase changes -  we also find examples of optimal transformations in which different costs are assigned to the same unitary transformation.
These problems can be traced back to the fact that the gauge symmetry resulting from changes in the phase of a quantum state is not taken into account in the complexity proposal, as already mentioned in \secref{sec:Gauge invariance}. Upon adding suitable boundary terms, gauge invariance is restored but the complexity vanishes.

Moreover, we consider examples of optimal transformations for different Kac-Moody groups. Here, in contrast to the Virasoro case, the complexity functional assigns non-zero cost to non-trivial transformations at least in some cases, after ensuring gauge invariance by suitable boundary terms.  We provide examples of such non-trivial optimal transformations and show that the cost and resulting target state are non-trivial as well.

\subsection{Virasoro group}
\label{sec:Ex_Vir}
For the Virasoro group, in \secref{sec:solutions eom Virasoro} we first derive solutions to the equations of motion of the complexity functional \eqref{eq:complexity_Virasoro_with_central_term} to obtain the associated optimal transformations. We then consider examples of optimal transformations belonging to the $U(1)$ and $SL(2,R)$ subgroup for the Virasoro group in \secref{sec:example_U(1)} and \ref{sec:example_SL(2,R)}. We calculate the complexities and target states associated with these transformations. Finally, in \secref{sec:example_SL(2,R)^n}, we consider a non-optimal transformation in the $SL(2,R)^n$ subgroup of the Virasoro group. We derive the associated complexity and compare the results with those obtained for optimal $SL(2,R)$ transformations.

\subsubsection{General solution to the equations of motion}
\label{sec:solutions eom Virasoro}
To derive the equations of motion of the complexity functional \eqref{eq:complexity_Virasoro_with_central_term}, we vary this functional with respect to $f$.
A straightforward but tedious calculation gives (see e.g.~\cite{Aldrovandi:1996sa})
\begin{equation}
  b_0\biggl(\frac{\dot f}{f'}\biggr)'-\frac{c}{48\pi}\biggl(\frac{\dot f}{f'}\biggr)''' = 0.
  \label{eq:eom_Virasoro} 
\end{equation}
Eq.~\eqref{eq:eom_Virasoro} is simple enough to be solved in the general case.
In fact, in terms of $(\dot f/f')'$ \eqref{eq:eom_Virasoro} is just the equation of motion for a harmonic oscillator with frequency $\omega_h = \sqrt{1-24h/c}$.
The solution is
\begin{equation}
  \frac{\dot f}{f'} = A(t)\sigma + B(t) e^{i\omega_h\sigma} + C(t) e^{-i\omega_h\sigma} + D(t),
  \label{eq:solution_eom_Virasoro1}
\end{equation}
where $A,B,C,D$ are arbitrary functions of $t$ satisfying $B(t) = C^*(t)$.
However, the requirement that $f$ be a diffeomorphism (at constant $t$) and hence that $\dot f$ and $f'$ be periodic demands that $A(t)=0$.
For $h > 0$, $\omega_h \notin \ZZ$ and thus in this case also $B(t) = C(t) = 0$.

We note that if $f(t,\sigma) = f_0(t,\sigma)$ is a solution of \eqref{eq:solution_eom_Virasoro1}, then
\begin{equation}
  f(t,\sigma) = g(f_0(t,\sigma))
  \label{eq:general solution eom Virasoro}
\end{equation}
is another solution of \eqref{eq:solution_eom_Virasoro1} if $g(\sigma)$ is a diffeomorphism, that is if $g' > 0$, $g(\sigma + 2\pi) = g(\sigma) + 2\pi$.
Therefore, it is sufficient to first determine  a simple solution $f_0(t,\sigma)$ of \eqref{eq:solution_eom_Virasoro1}.
The general solution then follows by composition with an arbitrary $g(\sigma)$.

Concentrate first on the case $h > 0$, where \eqref{eq:solution_eom_Virasoro1} reduces to
\begin{equation}
  \dot f = D(t) f'.
  \label{eq:solution_eom_Virasoro2}
\end{equation}
It is simple to see that
\begin{equation}
  f_0(t,\sigma) = \sigma + \delta(t)
  \label{eq:solution eom Virasoro h>0}
\end{equation}
provides a solution of this differential equation when $D(t) = \dot\delta(t)$.
Inserting the general solution $f(t,\sigma) = g(f_0(t,\sigma))$ into \eqref{eq:complexity_Virasoro_with_central_term}, we obtain that the complexity depends only on the value of $\delta(t)$ at $t=0$ and $t=T$,
\begin{equation}
  \CC = \int d\sigma dt\, \dot\delta(t)\biggl(b_0 + \frac{c}{48\pi}\(\frac{f''}{f'}\)'\biggr) = 2\pi b_0(\delta(T)-\delta(0)).
  \label{eq:complexity h>0}
\end{equation}

For $\delta(T) = \delta(0)$, the complexity vanishes.
As expected for a consistent interpretation, in this case the unitary transformation $U_f$ acting on the reference state is just the identity.
To see that, we note that $F(t,\sigma) = G(\sigma) - \delta(t)$ and thus $\epsilon(t,\sigma) = -\dot F/F' = \dot\delta(t)/G'(\sigma)$, where $G(g(\sigma)) = \sigma$.
Therefore, the time integral of $U_f=\cev{\CP}e^{\int\int dtd\sigma\,\epsilon(t,\sigma)T(\sigma)}$ is trivially solvable and $U_f$ is just an exponential of a sum over $L_n$ symmetry generators with prefactor $\delta(T)-\delta(0)$.
When $\delta(T) = \delta(0)$, this prefactor vanishes and the exponential yields  the identity.

For $h=0$, one solution to \eqref{eq:solution_eom_Virasoro1} is given by\footnote{An equivalent representation of this solution of the form
  \begin{equation}
    f(t,\sigma)=\frac{1}{i}\mathrm{ln}\Bigg[\frac{\alpha(t) e^{i\sigma}+\beta(t)}{\bar{\beta}(t)e^{i\sigma}+\bar{\alpha}(t)}\Bigg]
  \end{equation}
  was considered in \cite{Berry_phases}.}
\begin{equation}
  f_0(t,\sigma) = 2 \arctan\(\frac{a(t)\tan(\sigma/2)+b(t)}{c(t)\tan(\sigma/2)+d(t)}\),
  \label{eq:solution eom Virasoro h=0}
\end{equation}
where $ad-bc=1$, i.e.~$a,b,c,d$ together form a SL(2,R) element\footnote{PSL(2,R) to be precise since $a,b,c,d \to -a,-b,-c,-d$ gives the same $f_0$}.
The functions $a,b,c,d$ are related to $B,C,D$ by
\begin{equation}
  \begin{aligned}
    B &= \frac{1}{2}\biggl[\dot b d - b \dot d - \dot a c + a \dot c -i(\dot b c - b \dot c + \dot a d - a \dot d)\biggr] = C^*\\
    D &= \dot b d - b \dot d + \dot a c - a \dot c.
  \end{aligned}
\end{equation}
In this case, the complexity cannot be derived in closed form in general.
Inserting the general solution $f(t,\sigma) = g(f_0(t,\sigma))$ into the complexity action functional \eqref{eq:complexity_Virasoro_with_central_term} yields
\begin{align}
  \CC[f] & = \CC[f_0] + \frac{c}{48\pi} \int d\sigma dt\, \frac{\dot f_0}{f_0'}\(\frac{g''}{g'}\)' \nonumber\\
  & = \int dt D 2\pi b_0 + \frac{c}{48\pi} \int d\sigma dt\, (Be^{i\sigma}+Ce^{-i\sigma})\(\frac{f_0''}{f_0'}+\frac{g''}{g'}\)' \, .
  \label{eq:complexity h=0}
\end{align}

How do we interpret the above solutions \eqref{eq:solution eom Virasoro h>0} and \eqref{eq:solution eom Virasoro h=0} from the perspective of complexity?
We note that the  $f_0(t,\sigma)$ are in the stabilizer subgroup of the respective reference state, i.e.~$U_{f_0(t)}$ only generates a phase shift in $\ket h$.
Moreover, $g$ is independent of $t$.
The above solutions are given by a composition of $g$ with $f_0(t)$, hence
\begin{equation}
  U_{f(t)} = U_{g \cdot f_0(t)} \sim U_g U_{f_0(t)},
\end{equation}
where $\sim$ denotes equality up to a phase $e^{i\alpha}$.
Therefore, the state $\ket{\psi(t)}$ at computation time $t$ is reached by first applying a 
time-dependent phase shift $U_{f_0(t)}$ followed by a time independent conformal transformation $U_g$.
Thus, up to a phase, the target state is reached instantaneously at time $t=0$, \footnote{If one additionally requires that $U_{f(0)} = \mathbb{1}$ as in \cite{Caputa:2018kdj}, instantaneous transitions are forbidden. Then $g$ must be the identity and only phase shifts can be generated.}
\begin{equation}
  \ket{\psi_T} \sim U_{f(0)} \ket h \sim U_g \ket h.
\end{equation}
At later times $t>0$, only this phase changes.
What cost is associated with these two contributions to $U_{f(t)}$?
For $h>0$, from \eqref{eq:complexity h>0}, it is obvious that the complexity is independent of $g$ and counts only the phase change generated by $U_{f_0(t)}$.
Thus, it follows that in this case, the complexity definition given by the on-shell value of the geometric action \eqref{eq:complexity_Virasoro_with_central_term} counts only phase changes, while it assigns zero cost to the (most certainly more interesting) contribution that actually produces a physically distinct state!
For the special case of $h=0$, eq.~\eqref{eq:complexity h=0} shows that the complexity is in general not independent of $g$.
In this case, $U_g$ contributes to the cost of the total transformation $U_{f(t)}$ as does the phase shift generated by $U_{f_0(t)}$.

In the following, we derive the cost assigned to these phase shifts by considering solutions to the equations of motion where $g$ is the identity, i.e.~solutions for which only a phase change is generated.
These transformations for these solutions are generated by stabilizer subgroups U(1) and SL(2,R) of the Virasoro group.
We show that the cost is non-vanishing and proportional to the phase shift\footnote{Phase changes generated by conformal transformations, referred to as Virasoro Berry phases, were recently considered in \cite{Berry_phases}.
  See app.~\ref{app:Berry_phases} for a discussion how these Berry phases are related to complexity.}.

\subsubsection{Example: U(1) subgroup of the Virasoro group}
\label{sec:example_U(1)}
For $f(t,\sigma) = \sigma + \delta(t)$, only the $L_0$ generator forming a $U(1)$ subgroup of the Virasoro group acts on the reference states,
\begin{equation}
  \epsilon(t,\sigma) = -\dot F/F' \equiv \epsilon(t) = \dot\delta(t).
\end{equation}
For a reference state $\ket h$ with well-defined scaling dimension $h$, the complexity then counts only the change in phase as
\begin{equation}
  \ket h \to \cev{\CP}\exp\biggl(2\pi \int dt \epsilon(t) (L_0 - c/24)\biggr) \ket h.
  \label{eq:U(1) transformation}
\end{equation}
As we have already seen, the complexity is determined entirely by the boundary conditions for $\delta(t)$ at $t=0$ and $t=T$,
\begin{equation}
  \CC=\(h-\frac{c}{24}\)\(\delta(T)-\delta(0)\).
  \label{eq:U(1) complexity}
\end{equation}
Thus, despite the fact that we have only changed the phase of the state, which is a gauge symmetry, the complexity is still non-vanishing.

\subsubsection{Example: SL(2,R) subgroup of the Virasoro group}
\label{sec:example_SL(2,R)}
As a further example for evaluating the Virasoro complexity, we now consider optimal transformations belonging to the SL(2,R) subgroup of the Virasoro group. 
To this end, consider $f(t,\sigma)$ from \eqref{eq:solution eom Virasoro h=0}.
The inverse $F(t,\sigma)$ has the same form with $a,b,c,d$ replaced by $d,-b,-c,a$, which are the elements of the inverse of the SL(2,R) element corresponding to $a,b,c,d$.
Thus, $\epsilon(t,\sigma)$ is given by
\begin{equation}
  \epsilon(t,\sigma) = \tilde B(t) e^{i\sigma} + \tilde C(t) e^{-i\sigma} + \tilde D(t),
\end{equation}
where $\tilde B, \tilde C, \tilde D$ are functions of $t$ related to $a,b,c,d$ by
\begin{equation}
  \begin{aligned}
    \tilde B &= \frac{1}{2}\biggl[-\dot b a + b \dot a + \dot d c - d \dot c -i(\dot b c - b \dot c + \dot d a - d \dot a)\biggr] = {\tilde C}^*\\
    \tilde D &= -\dot b a + b \dot a - \dot d c + d \dot c.
  \end{aligned}
\end{equation}
Thus, only $L_0,L_{\pm 1}$ forming a SL(2,R) subgroup of the Virasoro group act on the reference state $\ket 0$,
\begin{equation}
  \ket{\psi_T} = U_{f(T,\sigma)}\ket{0}=\exp\(2\pi i \int_0^T dt \(\tilde B(t) L_{-1} + \tilde C(t) L_1 + \tilde D(t) \biggl(L_0-\frac{c}{24}\biggr)\)\)\ket{0}.
\end{equation}
Since $\ket 0$ is SL(2,R) invariant, this gives only a phase change that arises from the zero mode $L_0 - c/24$.
As for the $U(1)$ case, the complexity for these phase changing transformations does not vanish in general (see \eqref{eq:complexity h=0}).

Let us now consider a particular example of a SL(2,R) transformation for which the complexity does vanish.
We consider a diffeomorphism that was introduced in the context of Virasoro Berry phases in \cite{Berry_phases}.
There, $f(t,\sigma)$ is chosen such that the corresponding bulk diffeomorphism gives an $\mathrm{AdS_3}$ boost with rapidity $\lambda$,
\begin{equation}
f(t,\sigma)=\frac{1}{i}\mathrm{ln}\Bigg[\frac{\mathrm{cosh}\(\frac{\lambda}{2}\)e^{i\sigma}+\mathrm{sinh}\(\frac{\lambda}{2}\)}{\mathrm{sinh}\(\frac{\lambda}{2}\)e^{i\sigma}+\mathrm{cosh}\(\frac{\lambda}{2}\)}\Bigg]+\omega t.
	\label{eq:circular_path}
\end{equation}
The inverse transformation $F(t,\sigma)$ is a function of $\sigma - \omega t$, thus the only non-vanishing mode of $\epsilon(t)$ is the zero mode
\begin{align}
  \epsilon(t,\sigma) = \epsilon_0 = \omega \, .
\label{eq:modes}
\end{align}
The complexity follows from a straightforward calculation by inserting \eqref{eq:circular_path} into the action \eqref{eq:complexity_Virasoro_with_central_term},
\begin{equation}
	\CC=\CC_T=hT\omega \, \mathrm{cosh}\(\lambda\).
	\label{eq:complexity_SL(2,R)}
\end{equation}
For general $h$, this would imply that the cost induced by $\ket h \to \exp(i\int dt \, \epsilon_0(L_0-c/24))\ket h$ grows linearly with the angular velocity $\omega$ and computation time $T$.
However, for the vacuum we have $h=0$ such that  the complexity actually vanishes.
This is certainly a puzzling result in light of the fact that in \secref{sec:example_U(1)} we could have taken $\delta(t)=\omega t$ and derived a non-vanishing complexity \eqref{eq:U(1) complexity} for exactly the same unitary (phase changing) transformation \eqref{eq:U(1) transformation} as we have used here!
This discrepancy, which exists only for the vacuum state due to the fact that it is the only SL(2,R) invariant state, shows again that it is essential to take into account the gauge symmetry to obtain a consistent interpretation of geometric actions as complexity.

\subsubsection{Example: SL(2,R)$^n$ subgroup of the Virasoro group}
\label{sec:example_SL(2,R)^n}
Here, we consider transformations arising from the SL(2,R)$^n$ subgroup of the Virasoro group. Note that these do not solve the equations of motion \eqref{eq:eom_Virasoro} and are thus not optimal. It is interesting to compare the cost and target state with the results for optimal SL(2,R) transformations in sec.~\ref{sec:example_SL(2,R)}.

 To give an example for a non-optimal transformation, we generalize \eqref{eq:circular_path} to
\begin{equation}
  f(t,\sigma)=\frac{1}{in}\mathrm{ln}\Bigg[\frac{\alpha e^{in\sigma}+\beta}{\bar{\beta}e^{in\sigma}+\bar{\alpha}}\Bigg] + \omega t,
  \label{eq:circular_path_n}
\end{equation}
where $n \in \NN$.
Eq.~\eqref{eq:circular_path_n} is not a solution to the equations of motion \eqref{eq:eom_Virasoro}, due to the fact that $\dot f,f'$ contains terms proportional to $e^{\pm in\sigma}$.
The transformation arises from modes $L_0,L_{\pm n}$ forming a subgroup SL(2,R)$^n$ of the Virasoro group\footnote{This is the stabilizer group of so-called exceptional coadjoint orbits of the Virasoro group. For $n>1$, these orbits are unphysical with energy unbounded from below \cite{BMS}, thus they are not considered in the rest of this work.}.
SL(2,R)$^n$ is an $n$-fold cover of SL(2,R).
The associated cost is non-minimal and not a measure for complexity, as we will show now.

Since the inverse transformation $F(t,\sigma)$ is a function of $\sigma - \omega t$, the only non-vanishing mode for this transformation is again the zero mode
\begin{align}
  \epsilon(t,\sigma) = \epsilon_0 = \omega.
\label{eq:modes_n}
\end{align}
Thus, the unitary operator $U_f$ implementing the transformation $f(t,\sigma)$ in the Hilbert space generates the same state as the SL(2,R) transformation considered above.
The cost of this transformation is given by
\begin{equation}
  \CC = hT\omega \mathrm{cosh}(\lambda) + \omega T\frac{c}{24}\(n^2-1\)\mathrm{cosh}(\lambda).
  \label{eq:cost SL(2,R)^n}
\end{equation}
As expected, the complexity of non-optimal transformations in the $n$-fold cover is larger than that of optimal SL(2,R)-transformations in eq.~\eqref{eq:complexity_SL(2,R)}.
However, the corresponding unitary operator $U_f$ acting on the Hilbert space still contains only the $L_0$ generator and is exactly equivalent to the one for the (optimal) SL(2,R) path of sec. \ref{sec:example_SL(2,R)}.
Thus we see again that without taking the gauge invariance into account, the complexity definition assigns different costs \eqref{eq:cost SL(2,R)^n} and \eqref{eq:complexity_SL(2,R)} to the same unitary transformation $U_f$.

\subsection{Kac-Moody groups}
\label{sec:Ex_Kac}
We now discuss complexity for Kac-Moody groups for the examples of simple optimal $SU(2)$, $SL(2,R)$, and $SL(N,R)$ transformations. We begin by deriving the equation of motion and determining a general solution in sec.~\ref{sec:general_remarks}. There, we also show that the first term in the complexity \eqref{eq:complexity_Kac-Moody} can never contribute when group elements are represented by unit determinant matrices. We conclude sec.~\ref{sec:general_remarks} by discussing relevant aspects for the explicit computation of the topological WZW term.

In sec.~\ref{sec:example_SU(2)}, we then proceed to compute the complexity of $SU(2)$ symmetry transformations. The  optimal transformations we consider show that in contrast to the Virasoro case, the complexity measure \eqref{eq:complexity_Kac-Moody} may assign non-vanishing cost to non-trivial transformations, as expected for a viable complexity measure. Moreover, since we argued in sec. \ref{sec:Gauge invariance} that the complexity is gauge invariant only up to boundary terms, we demonstrate for an example that similarly to the Virasoro case, the complexity proposal considered assigns also a cost to gauge transformations, unless appropriate boundary terms are added to the complexity functional \eqref{eq:complexity_Kac-Moody}. In sec.~\ref{sec:example_KM_SL2}, we then provide a further example of an optimal $SL(2,R)$ transformation yielding non-vanishing complexity for a non-trivial target state. Finally, in sec.~\ref{sec:example_SLn} we  study how the complexity cost assigned to gauge transformations grows with the value of $N$ for $SL(N,R)$ for the example of diagonal matrices. 

\subsubsection{Optimal circuits for Kac-Moody complexity}
\label{sec:general_remarks}
To derive the equations of motion of \eqref{eq:complexity_Kac-Moody}, we note that the variation of \eqref{eq:complexity_Kac-Moody} coincides with that of a WZW model (see e.g. \cite{Witten_WZW}) since the term scaling with $v_0$ in \eqref{eq:complexity_Kac-Moody} does not contribute.
The equation of motion is then that of a WZW model,
\begin{equation}
  0=\partial_t(\partial_\sigma\Omega\Omega^{-1}).
  \label{eq:eom_Kac-Moody}
\end{equation}
Here, $\Omega$ are matrices valued in the semisimple Lie group specifying the Kac-Moody group as introduced in sec. \ref{sec:complexity_Kac-Moody}.
The relation of \eqref{eq:eom_Kac-Moody} to the equations of motion of the WZW model becomes apparent if we identify $t$ with a holomorphic coordinate, $\sigma$ with an antiholomorphic coordinate and set $\partial_{\sigma }\Omega\Omega^{-1}=J_\sigma$. Then the equation of motion \eqref{eq:eom_Kac-Moody} becomes equivalent to the conservation equation $\partial_{z}J_{\bar{z}}=0$. 

General solutions of \eqref{eq:eom_Kac-Moody} are given by solutions of the WZW model, which read \cite{Witten_WZW}
\begin{equation}
\Omega(t,\sigma)=\Omega_{1}(\sigma)\Omega_2(t).
\label{eq:general_solution}
\end{equation}
Next, let us comment on the special role of the orbit label $v_0$ in the Kac-Moody case.
Note that the orbit $v_0$ does not contribute to the equation of motion \eqref{eq:eom_Kac-Moody} for the Kac-Moody complexity. This distinguishes the equation of motion from those obtained for the Virasoro case \eqref{eq:eom_Virasoro}. Moreover, the term scaling with $v_0$ in \eqref{eq:complexity_Kac-Moody} vanishes for any unit-determinant matrix and thus does not contribute to the complexity at all for these. This may be shown with Jacobi's formula, which for an invertible matrix $A$ reads
\begin{equation}
\partial\mathrm{det}[A]=\mathrm{det}[A]\tr[\partial A A^{-1}].
\end{equation}
Setting $\mathrm{det}[A]=1$, we hence obtain
\begin{equation}
0	= \tr[\partial A A^{-1}].
\end{equation}
 The right-hand side may now be identified with $\tr[\dot{\Omega}\Omega^{-1}]$ in \eqref{eq:complexity_Kac-Moody} upon realizing that $v_0=-\frac{1}{2\pi}C_\lambda\propto\mathbf{1}$ in an irreducible representation according to Schur's lemma. Consequently, the complexity is independent of the expectation value of $J$ and thus the Casimir element $C_\lambda$ for all transformations that may be written as a matrix with unit determinant. This implies that the complexity is independent of the specific coadjoint orbit, which is in stark contrast to the Virasoro complexity \eqref{eq:complexity_Virasoro_with_central_term}, where even upon considering the simplest transformations, terms scaling with the expectation value of the energy-momentum tensor never vanish. Indeed, the only conceivable way to mimic this behavior in the Virasoro case is to consider those diffeomorphisms that do not depend on time such that $\dot{f}$ vanishes. This, however, always leads to a vanishing complexity. 
 We will further illustrate the consequences of the vanishing term scaling with the Casimir element on the example of $SU(2)$ transformations in sec.~\ref{sec:example_SU(2)}.
 
Next, let us take a closer look at the topological WZW term, which is the last term in \eqref{eq:complexity_Kac-Moody}. We briefly review aspects relevant for its explicit calculation for examples of optimal transformations considered below. The 2d CFT WZW term is discussed in \cite{Witten_WZW}, a more detailed discussion of its construction and properties may be found in \cite{Witten_WZW_general}. 

Rewritten in a more intuitive notation, the topological WZW term introduced in \eqref{eq:geometric_action} reads
\begin{equation}
\Gamma_{WZW}=-\frac{k}{24\pi}\int_0^1 dr \int_{\Sigma} d^2x\, \tr\left[ \epsilon_{\alpha\beta\gamma}\partial_\alpha\widetilde{\Omega}\widetilde{\Omega}^{-1}\partial_\beta\widetilde{\Omega}\widetilde{\Omega}^{-1}\partial_\gamma\widetilde{\Omega}\widetilde{\Omega}^{-1}\right].
\label{eq:WZW}
\end{equation}
As briefly mentioned in sec.~\ref{sec:Geometric_Action_Kac-Moody}, the integral is now over a three-dimensional manifold $M=[0,1]\times\Sigma$, which is obtained by first compactifying the original two-dimensional manifold spanned by $t$ and $\sigma$ and parametrizing the interior of this compactified space by $r\in[0,1]$. The three-dimensional manifold thus must be chosen such that its boundary is the compactified version of the manifold spanned by $t$ and $\sigma$, $\partial M=\Sigma$. The new coordinates for the compactified space $\Sigma$ are denoted by $x$ in \eqref{eq:WZW}. Accordingly, the transformations $\Omega(t,\sigma)$ have to be extended into the interior. The extended version is denoted by $\widetilde{\Omega}$ in \eqref{eq:WZW}. The existence of such extensions is guaranteed if the 2nd homotopy group of the  Lie group $G$ vanishes, i.e. $\pi_2(G)=0$. However, the particular way in which the map is extended is not unique. In order for $\widetilde{\Omega}(r,x): [0,1]\times \Sigma\rightarrow G$ to be a homotopy, the extension simply has to obey the boundary conditions\footnote{To be precise, the extension simply has to satisfy the more general conditions $\widetilde{\Omega}(r=0,x)=\Omega_1(x)$ and  $\widetilde{\Omega}(r=1,x)=\Omega_2(x)$. The condition $\widetilde{\Omega}(r=0,x)=\mathbf{1}$ is a more specific but convenient choice. For a pedagogical introduction into homotopies, we refer to ch. 4 in \cite{Nakahara:2003nw}, their relation to the WZW term in discussed in \cite{Witten_WZW}, \cite{Witten_WZW_general}, as well as \cite{Dobado:1997jx} ch. 4.10. } 
\begin{align}
\widetilde{\Omega}(r=0,x)=\mathbf{1} \qquad \widetilde{\Omega}(r=1,x)=\Omega(x).
\end{align} 
Note that the second condition is equivalent to requiring that we obtain the original transformation at the boundary of the manifold $M$.
One possibility to extend the transformation is\footnote{The result does not depend on the particular choice of $\widetilde{\Omega}$.} \cite{Fujiwara_2003}
\begin{equation}
\widetilde{\Omega}(x,r)=\left(\Omega(x)\right)^r.
\label{eq:extension}
\end{equation}
This will be used to calculate the contribution of the topological WZW term \eqref{eq:WZW} to the Kac-Moody complexity for some examples of optimal transformations for the $SU(2)$, $SL(2,R)$, and $SL(N,R)$ Kac-Moody groups.

\subsubsection{Example: SU(2) Kac-Moody group}
\label{sec:example_SU(2)}
We have observed in sec. \ref{sec:Ex_Vir} for the Virasoro case that the geometric action \eqref{eq:complexity_Virasoro_with_central_term} as complexity measure assigns vanishing cost to any non-trivial transformation for all reference states except the vacuum. At the same time, it assigns a non-trivial  cost to gauge transformations that yield only phase changes on the reference state. We explained in sec. \ref{sec:Gauge invariance} that the non-trivial results for gauge transformations are due to a lack of gauge invariance of the geometric action that occurs when the geometric action is given in terms of a group path rather than a path on the coadjoint orbit. 

In the Kac-Moody case however, the situation is somewhat different. With the examples considered here, we aim at demonstrating two properties of the Kac-Moody complexity measure \eqref{eq:complexity_Kac-Moody}: First, it counts the cost of non-trivial transformations that lead to target states physically distinguishable from the reference state. This is an essential difference to the Virasoro case examined in sec. \ref{sec:Ex_Vir}. Secondly, since the Kac-Moody geometric action, which is our complexity measure \eqref{eq:complexity_Kac-Moody}, is defined in terms of a group path in the group manifold rather than on the coadjoint orbit, it also counts the cost of trivial gauge transformations as in the Virasoro case. These may however be removed by appropriate boundary terms. We also examine further general properties of the complexity measure \eqref{eq:complexity_Kac-Moody}. To be specific, we choose optimal SU(2) transformations.  However, the general results apply to other groups as well.

We begin with an example for a transformation yielding a non-trivial target state.
In the fundamental representation, the generators of SU(2) are given in terms of the Pauli matrices $\sigma$,
\begin{align}
\sigma_1=
\begin{pmatrix}
0&1\\1&0
\end{pmatrix},
\qquad
\sigma_2=
\begin{pmatrix}
0&-i\\i&0
\end{pmatrix},
\qquad\sigma_3=
\begin{pmatrix}
1&0\\0&-1
\end{pmatrix}.
\label{eq:Pauli}
\end{align}
A non-trivial transformation solving the equation of motion \eqref{eq:general_solution} is given by
\begin{equation}
\Omega(t,\sigma)=
\begin{pmatrix}
\mathrm{cos}(\sigma)& \mathrm{sin}(\sigma)\\
\mathrm{-sin}(\sigma)&\mathrm{cos}(\sigma)
\end{pmatrix}
\begin{pmatrix}
\mathrm{cos}( t)& \mathrm{sin}( t)\\
\mathrm{-sin}( t)&\mathrm{cos} (t)
\end{pmatrix},
\label{eq:transformartion_SU(2)}
\end{equation}
which is a sequence of rotations generated by $\sigma_2$. To evaluate the topological term, we apply the general extension \eqref{eq:extension} to the particular solution considered here.
Inserting the optimal transformation \eqref{eq:transformartion_SU(2)} into \eqref{eq:complexity_Kac-Moody} yields
\begin{equation}
	\CC=kT
	\label{eq:complexity_rotation}.
\end{equation}
As anticipated in sec.~\ref{sec:general_remarks}, the value of $C_\lambda$ is irrelevant as the first term in \eqref{eq:complexity_Kac-Moody}, which scales with $v_0$ and thus $C_\lambda$, vanishes. In particular, $C_j=2j(j+1)$ in a spin-$j$ representation of $SU(2)$. The result \eqref{eq:complexity_rotation} then implies the complexity is independent of the spin for any $SU(2)$ transformation. Moreover, the topological term does not contribute. This is a peculiarity of the generators involved in the transformation. The transformation \eqref{eq:transformartion_SU(2)} may be written in terms of $\mathbb{1}$ and $\sigma_2$, which commute. The fact that the integrand of the topological term is antisymmetric requires certains terms to cancel  in the complexity. 

In place of \eqref{eq:transformartion_SU(2)}, we may also consider the more general non-trivial transformation
\begin{equation}
		\Omega=
	\begin{pmatrix}
	\mathrm{cos}(\alpha\sigma+\delta)& \mathrm{sin}(\alpha\sigma+\delta)\\
	\mathrm{-sin}(\alpha\sigma+\delta)&\mathrm{cos}(\alpha\sigma+\delta)
	\end{pmatrix}
	\begin{pmatrix}
	\mathrm{cos}(\beta t+\lambda)& \mathrm{sin}(\beta t+\lambda)\\
	\mathrm{-sin}(\beta t+\lambda)&\mathrm{cos}(\beta t+\lambda)
	\end{pmatrix},
	\label{eq:trafo_non_trivial_state}
\end{equation}
where $\alpha, \beta, \delta, \lambda \in \mathbb{R}$.
With \eqref{eq:complexity_Kac-Moody}, we obtain
\begin{equation}
	\CC=\alpha\beta kT,
	\label{eq:example_SU(2)_scaling}
\end{equation}
where the only contribution again arises from the $2$nd term in \eqref{eq:complexity_Kac-Moody}. Apparently, a scaling of the coordinates by a real number leads to the same scaling of the complexity with $\alpha$. This behavior is caused by the derivatives in the $2$nd term of the complexity. In contrast, adding $\alpha$ to $\sigma$ or $t$ does not change the complexity.
A simple consistency check consists of setting $\alpha=\beta=1$ in \eqref{eq:example_SU(2)_scaling}, which yields \eqref{eq:complexity_rotation}.

We have thus demonstrated that the transformation \eqref{eq:trafo_non_trivial_state} yields a non-vanishing complexity \eqref{eq:example_SU(2)_scaling}. Since we aim to show that we obtain non-vanishing cost for non-trivial transformations, we now determine the target state associated with the transformation \eqref{eq:example_SU(2)_scaling}. As a reference state, we chose a highest weight state given by $\ket{h,j_{\mathrm{max}}}$, where $h$ denotes the conformal weight and $j_{\mathrm{max}}$ denotes the maximum spin, i.e. the reference state is a highest weight state of the zero-mode group $SU(2)$ as well. To obtain the target state, we employ 
\begin{equation}
	\ket{\psi_T}=\cev{\CP}e^{\int_0^T dt\int_0^{2\pi}\frac{d\sigma}{2\pi} \tr[\epsilon(t,\sigma)J(\sigma)]}\ket{h,j_{\mathrm{max}}}=\cev{\CP}e^{\int_0^T dt\,\tr[\epsilon^{a}_n(t)J^{a}_{-n}]}\ket{h,j_{\mathrm{max}}}.
	\label{eq:target_state_KM}
\end{equation}
The index $a$ runs over the three generators of the group $SU(2)$ given in terms of the Pauli matrices by $J^a=\frac{\sigma^a}{\sqrt{2}}$, whereas $n\in \mathbb{Z}$ as for the Virasoro case. The velocity modes $\epsilon^a(t,\sigma)$ are given by 
\begin{equation}
	\epsilon^{a}(t, \sigma)=\tr\left[\epsilon(t, \sigma)\frac{\sigma^a}{\sqrt{2}}\right]=\tr\left[\partial_t\Omega(t,\sigma)\Omega^{-1}(t,\sigma)\frac{\sigma^a}{\sqrt{2}}\right],
	\label{eq:modes_KM}
\end{equation}
where in the last step we used \eqref{eq:velocity_Kac-Moody}. Evaluating this yields $\epsilon^2=\sqrt{2}i\beta$ with $\epsilon^1=\epsilon^3=0$. Since $\epsilon^2=\mathrm{const}$, only the zeroth Fourier mode is non-vanishing. Hence, $\epsilon^2_0=\sqrt{2}i\beta$ with all other modes vanishing. For the target state  this implies
\begin{equation}
	\ket{\psi_T}=e^{i\sqrt{2}\beta T\,J^{2}_{0}}\ket{h,j_{\mathrm{max}}}.
\end{equation}
To proceed,
$J^2_0$ can be expressed in terms of the ladder operators $J^{\pm}$. The target state may then be obtained by expanding the exponential and evaluating the action of the operators $J^{\pm}$ on the reference state. Since the ladder operators change the value of $j_\mathrm{max}$, the target state is a sum of states with different values of $j_\mathrm{max}$. They, however, do not change the conformal dimension $h$. Hence, the target state is still a primary state.  Nevertheless, we see an important feature of the target state already at this point. It is non-trivial as the different values of $j_\mathrm{max}$ make it distinguishable from the reference sate. Therefore, in the Kac-Moody case, there exist optimal circuits with non-vanishing cost that lead to physically distinguishable target states. Here, the non-trivial transformation \eqref{eq:trafo_non_trivial_state} has non-vanishing complexity \eqref{eq:example_SU(2)_scaling} and yields a sum of conformal primary target state $\ket{h,j}$ with different values of $j$. This is an important difference to the Virasoro case, where for any reference state with $h>0$, the complexity only counts the cost of phase changes as discussed in sec. \ref{sec:Ex_Vir}.

However, also in the Kac-Moody case, there are examples where the process described counts the cost of gauge transformations. This is due to the fact that the geometric action \eqref{eq:complexity_Kac-Moody} is gauge invariant only up to boundary terms if it is defined in terms of a group path. The coadjoint orbits for Kac-Moody groups $\hat{LG}$ are given by $\hat{LG}/H_{v_0}$. As discussed in sec. \ref{sec:Gauge invariance}, $H_{v_0}$ is the stabilizer subgroup that leaves elements on the orbit invariant. For Kac-Moody groups based on compact subgroups, the stabilizer subgroup $H_{v_0}$ is given by the maximal torus, i.e. the maximal compact Abelian subgroup  \cite{pressley1988loop}. Therefore, to demonstrate that gauge transformations lead to non-vanishing cost even though the target state differs only by a phase change from the reference state, we first pick an optimal transformation from the orbit stabilizer and then compute the complexity and the target state. 
For $SU(2)$, an example of such a gauge transformation is given by
\begin{equation}
\Omega=
\begin{pmatrix}
e^{i\sigma}	& 0\\
0	& e^{-i\sigma}
\end{pmatrix}
\begin{pmatrix}
e^{i2\pi t}& 0\\
0	& e^{-i2\pi t}
\end{pmatrix}.
\label{eq:stabilizer}
\end{equation}
Inserting the transformation into \eqref{eq:complexity_Kac-Moody} yields
\begin{equation}
\CC=2\pi kT, \label{pikT}
\end{equation}
which is evidently non-vanishing, unlike what we expect for a gauge transformation since the reference state only changes by a phase. Indeed, the associated target is given by
\begin{equation}
\ket{\psi_T}=e^{i\frac{4}{\sqrt{2}}\pi T\,J^{3}_{0}}\ket{h,j_{\mathrm{max}}} \, ,
\end{equation}
as only the velocity mode $\epsilon_0^3$ is not equal to zero.
With $J_0^3$ related to the ladder operator convention by $J^z_0=\frac{1}{\sqrt{2}}J_0^3$ and  $J_0^z\ket{h,j_{\mathrm{max}}}=\frac{j_{\mathrm{max}}}{2}\ket{h,j_{\mathrm{max}}}$,
we then obtain
\begin{equation}
\ket{\psi_T}=e^{2i\pi Tj_{\mathrm{max}}}\ket{h,j_{\mathrm{max}}}.
\end{equation}
This is the reference state up to a phase, but the complexity \eqref{pikT} is non-vanishing.

As discussed in sec. \ref{sec:Ex_Vir}, we expect that \eqref{pikT} will be set to zero by
 appropriate boundary terms, which we now derive.
 We may write a general transformation from the orbit stabilizer as
 \begin{equation}
 	\Omega=
 	\begin{pmatrix}
 	e^{i\alpha(\sigma)}	& 0\\
 	0	& e^{-i\alpha(\sigma)}
 	\end{pmatrix}
 	\begin{pmatrix}
 	e^{i \beta(t)}& 0\\
 	0	& e^{-i\beta( t)}
 	\end{pmatrix}.
 	\label{eq:stabilizer_general}
 \end{equation}
As discussed in sec. \ref{sec:general_remarks}, the first term in \eqref{eq:complexity_Kac-Moody} vanishes for all unit-determinant groups. Moreover, the WZW term, which is the last term in \eqref{eq:complexity_Kac-Moody}, is antisymmetric and thus vanishes for abelian subgroups. The boundary term cancelling the contribution \eqref{pikT} must then arise from the second term in \eqref{eq:complexity_Kac-Moody}. Inserting \eqref{eq:stabilizer_general} into \eqref{eq:complexity_Kac-Moody} thus yields the boundary term
\begin{equation}
	\CC_\mathrm{boundary}=-\frac{k}{2\pi}\int_0^T dt\int_0^{2\pi} d\sigma\,\partial_t \beta(t)\partial_{\sigma }\alpha(\sigma)=-\frac{k}{2\pi}(\beta(T)-\beta(0))(\alpha(2\pi)-\alpha(0)).
	\label{eq:boundary_KM}
\end{equation}
It is easy to check that \eqref{eq:boundary_KM} indeed cancels the complexity \eqref{pikT}. For the transformation \eqref{eq:stabilizer}, $\alpha(\sigma)=\sigma$ and $\beta(t)=2\pi t$. Inserting this into \eqref{eq:boundary_KM}, we obtain
\begin{equation}
		\CC_\mathrm{boundary}=-k2\pi T.
\end{equation}
This exactly cancels the complexity \eqref{pikT} and shows that gauge invariance may again be restored by an appropriate boundary term, which is given by \eqref{eq:boundary_KM} for group paths from the maximal torus.

We have therefore demonstrated that the complexity measure \eqref{eq:complexity_Kac-Moody} assigns non-vanishing cost to both trivial and non-trivial transformations. The cost of trivial transformations from the orbit stabilizer may however be cancelled by appropriate boundary terms. Note that above we examined examples where the transformation is either trivial or non-trivial, it is however also possible to obtain optimal transformations that contain  both a non-trivial and a trivial part. This is similar to the Virasoro case examined in sec. \ref{sec:Ex_Vir}. However, for the Virasoro group, this non-trivial part has zero cost for all reference states except the vacuum. For the Kac-Moody case however, we found examples for non-trivial transformations with non-zero cost, as is desidered for a meaningful complexity definition.  The reason for this is most easily understood by considering the target state \eqref{eq:target_state_KM}. The velocity modes $\epsilon^a_n$ determine which generators act on the reference state and thus determine the target state. The velocity $\epsilon(t,\sigma)$ is in turn given by the optimal transformation according to \eqref{eq:velocity_Kac-Moody}. We may distinguish several cases. For the examples considered above, $\epsilon(t,\sigma)=\mathrm{const}$, which implies that only $n=0$ modes do not vanish. Therefore, only the $SU(2)$ subgroup generators, i.e. the zero-mode generators of the Kac-Moody group contribute in the exponent of \eqref{eq:target_state_KM}. The general target state $\eqref{eq:target_state_KM}$ for $\epsilon(t,\sigma)=\mathrm{const}$ then reads
\begin{equation}
	\ket{\psi_T}=e^{iT\sum_a \epsilon_0^aJ_0^a}\ket{h,j_\mathrm{max}},
	\label{eq:target_state_n0}
\end{equation}
where $J_0^a$ are the $SU(2)$ subalgebra generators. The modes $\epsilon^a_0$ that do not vanish determine which of these generators act on the reference state. For instance,  if $\epsilon^1$ and $\epsilon^2$ do not vanish,  the target state must be non-trivial since the $SU(2)$ generators $J^1$ and $J^2$ may be written in terms of ladder operators, which act non-trivially on the state. An example of such a transformation is \eqref{eq:trafo_non_trivial_state} with non-vanishing complexity \eqref{eq:example_SU(2)_scaling}. On the other hand, if the mode $\epsilon^3$ is non-vanishing, the resulting target state will always contain a phase change even if $\epsilon^1$ and $\epsilon^2$  are non-vanishing as well. The reason is simple: The reference state is an eigenstate to the generator $J^3$ that acts on the reference state if the associated mode $\epsilon^3$ is non-vanishing. We may then substitute $J^3$ in \eqref{eq:target_state_n0} with its eigenvalue. Then, $\epsilon^3$ is multiplied by the eigenvalue in \eqref{eq:target_state_n0}, which is just a phase. An example of such a transformation is \eqref{eq:stabilizer}. Note that while for \eqref{eq:stabilizer} only $\epsilon^3$ is non-vanishing, it is in general conceivable to find transformations where also $\epsilon^1$ and $\epsilon^2$ are non-vanishing. Then, $\epsilon^3$ will still yield a phase since the reference state is an eigenstate to $J^3$, while $\epsilon^1$ and $\epsilon^2$ and the associated ladder operators transform the state in a non-trivial manner. Of course, there also exits optimal transformations where not only the zero modes $\epsilon^a_0$ but other modes $\epsilon^a_n$ contribute. This case can often be obtained by considering asymmetric transformations, i.e. where the matrices $\Omega_1(\sigma)$ and $\Omega_2(t)$ in \eqref{eq:general_solution} are different matrices from the same group. In this case, the Kac-Moody generators $J_n$ also transform the state non-trivially by changing the conformal dimension $h$. Nevertheless, if the optimal transformation yields a non-vanishing mode $\epsilon_0^{3}$, the target state will contain a phase change whose cost is cancelled by appropriate boundary terms. In general, any complexity contribution associated to transformations from the orbit stabilizer can be set to zero by boundary terms, as such group paths are projected onto points on the orbit.

\subsubsection{Example: SL(2,R) Kac-Moody group} 
\label{sec:example_KM_SL2}
To further demonstrate that the Kac-Moody complexity functional \eqref{eq:complexity_Kac-Moody} assigns cost to non-trivial transformations, we consider an optimal non-trivial $SL(2,R)$ transformation.
A general $SL(2,R)$ transformation may be written as
\begin{equation}
	\Omega=
\begin{pmatrix}
	a	& b\\
	c	& d
\end{pmatrix},
\label{eq:general_SL}
\end{equation}
where $ab-cb=1$ and $a,b,c,d \in \mathbb{R}$. Here, $a,b,c,d$ will be some functions of $\sigma$ and $t$. The generators can be expressed in terms of the Pauli matrices \cite{Maldacena_AdS_SL2R},
\begin{align}
	J_0=-\frac{i}{2}\sigma_2, \qquad J_{\pm}=\frac{1}{2}(\sigma_3\pm i\sigma_1).
	\label{eq:generators_SU(2)}
\end{align}
 Now consider the non-trivial optimal transformation
\begin{equation}
	\Omega=
\begin{pmatrix}
	\mathrm{cosh}(\alpha\sigma+\delta)& \mathrm{sinh}(\alpha\sigma+\delta)\\
	\mathrm{sinh}(\alpha\sigma+\delta)&\mathrm{cosh}(\alpha\sigma+\delta)
\end{pmatrix}
\begin{pmatrix}
\mathrm{cosh}(\beta t+\lambda)& \mathrm{sinh}(\beta t+\lambda)\\
\mathrm{sinh}(\beta t+\lambda)&\mathrm{cosh}(\alpha t+\lambda)
\end{pmatrix}.
\label{eq:non-trivial_SL}
\end{equation} 
We now show that the associated target state and complexity are both non-trivial. We begin with computing the complexity. Inserting the transformation \eqref{eq:non-trivial_SL} into the complexity measure \eqref{eq:complexity_Kac-Moody} yields
\begin{equation}
	\CC=\alpha\beta kT.
	\label{eq:complexity_non-trivial_SL2}
\end{equation}
Note that this result is identical to that of $SU(2)$ in \eqref{eq:example_SU(2)_scaling}. In particular, the derivatives in \eqref{eq:complexity_Kac-Moody} lead to the same scaling behavior. Moreover, just as observed for $SU(2)$, there is no contribution from the topological term. This vanishes for similar reasons as discussed in sec.~\ref{sec:example_SU(2)}, since the generators of $SL(2,R)$ may be constructed from Pauli matrices as given in \eqref{eq:generators_SU(2)}. 

Next, we show that similarly to the $SU(2)$ case considered in sec. \ref{sec:example_SU(2)}, the target state of the $SL(2,R)$ transformation \eqref{eq:non-trivial_SL} is non-trivial. An analogous calculation to that in sec.~\ref{sec:example_SU(2)} shows that the mode $\epsilon^1=\beta$ associated with the generator $J_1=\frac{1}{2}\sigma_1$ is non-vanishing. Since according to \eqref{eq:generators_SU(2)}, $J_1=-\frac{i}{2}(J_+-J_-)$, the target state is then given by
\begin{equation}
	\ket{\psi_T}=e^{-\frac{i}{2}\beta(J_+-J_-)}\ket{\psi_R}.
\end{equation}  
The ladder operators act non-trivially on the reference state $\ket{\psi_R}$. Therefore, the resulting target state is non-trivial. We hence observe that similarly to the optimal SU(2) transformation \eqref{eq:example_SU(2)_scaling}, there exist optimal transformations that yield target states which are physically distinguishable from the reference state. For SL(2,R), an example of such an optimal transformation is given by \eqref{eq:non-trivial_SL}. Most importantly, in contrast to the Virasoro group, the complexity measure \eqref{eq:complexity_Kac-Moody} assigns cost to these non-trivial transformations as we obtain a non-vanishing complexity \eqref{eq:complexity_non-trivial_SL2} for SL(2,R) and \eqref{eq:example_SU(2)_scaling} for SU(2).

\subsubsection{Example: diagonal matrices from the SL(N,R)  Kac-Moody group}
\label{sec:example_SLn}
Another simple solution of \eqref{eq:general_solution} are diagonal matrices, which hold the advantage that the computation can be easily extended to $SL(N,R)$. Moreover, the topological WZW term, which is the third term in the complexity functional \eqref{eq:complexity_Kac-Moody}, does not contribute since diagonal matrices commute and thus all contributions cancel due to the antisymmetry of the integrand. While these abelian matrices are gauge transformations yielding trivial target states, the non-vanishing complexity we obtain gives us some insight into how the complexity changes for a general $SL(N,R)$ group. 

 Let us begin by considering diagonal $SL(2,R)$ matrices. A general solution to the equation of motion \eqref{eq:general_solution} is then given by 
\begin{equation}
\Omega(t,\sigma)=
\begin{pmatrix}
\frac{1}{a(\sigma)}&0\\
0&a(\sigma)
\end{pmatrix}
\begin{pmatrix}
\frac{1}{b(t)}&0\\
0&b(t)
\end{pmatrix} \, ,
\end{equation}
with arbitrary but real-valued functions $\alpha(\sigma )$ and $b(t)$.
Inserting these solutions into \eqref{eq:complexity_Kac-Moody} again leads to a functional of the form
\begin{equation}
	\CC=\frac{k}{2\pi}\int_0^T\int_0^{2\pi}dtd\sigma\,\frac{a^{\prime}(\sigma)\dot{b}(t)}{ a(\sigma)b(t)}.
	\label{eq:complexity_diagonal_2x2}
\end{equation}
 Similarly, an according $SL(3,R)$ matrix reads 
\begin{equation}
	\Omega(t,\sigma)=
\begin{pmatrix}
	1/a^2(\sigma)&0&0\\
	0&a(\sigma)&0\\
	0&0&a(\sigma)
\end{pmatrix}
\begin{pmatrix}
1/b^2(t)&0&0\\
0&b(t)&0\\
0&0&b(t)
\end{pmatrix}.
\end{equation}
From \eqref{eq:complexity_Kac-Moody}, we obtain
\begin{equation}
	\CC=\frac{3k}{2\pi}\int_0^T\int_0^{2\pi}dtd\sigma\,\frac{a^{\prime}(\sigma)\dot{b}(t)}{ a(\sigma)b(t)}.
\end{equation}

A comparison with \eqref{eq:complexity_diagonal_2x2} leads to the conclusion that only prefactors change. This allows us to deduce the complexity of a matrix $n\times n \in SL(N,R)$,
\begin{equation}
\Omega(t,\sigma)=
\begin{pmatrix}
\frac{1}{a^{n-1}(\sigma)}&0&\cdots&0\\
0&a(\sigma)&\ddots&\vdots\\
\vdots & \ddots&\ddots&0\\
0&\cdots&0&a(\sigma)
\end{pmatrix}
\begin{pmatrix}
\frac{1}{b^{n-1}(t)}&0&\cdots&0\\
0&b(t)&\ddots&\vdots\\
\vdots & \ddots&\ddots&0\\
0&\cdots&0&b(t)
\end{pmatrix}.
\end{equation}
Inserting this into \eqref{eq:complexity_Kac-Moody} yields
\begin{equation}
	\CC=\frac{n(n-1)k}{4\pi}\int_0^T\int_0^{2\pi}dtd\sigma\,\frac{a^{\prime}(\sigma)\dot{b}(t)}{ a(\sigma)b(t)}.
\end{equation}
While these results seem promising that complexity may be obtained for arbitrary $n\times n$ matrices, an important aspect to keep in mind is that both the complexity functionals \eqref{eq:complexity_Virasoro_with_central_term} and \eqref{eq:complexity_Kac-Moody} count cost of trivial transformations such as gauge transformations. Note, however, the important difference to the Virasoro case: The restrictions on the solutions are far more general in the Kac-Moody case as any product of a $t$- and $\sigma$-dependent symmetry transformation solves \eqref{eq:eom_Kac-Moody}. Therefore, even upon subtracting appropriate terms arising from the orbit stabilizer, there exist non-trivial solutions with non-vanishing complexity that produce distinguishable target states. Two simple examples are given by \eqref{eq:trafo_non_trivial_state} and \eqref{eq:non-trivial_SL}, leading to the non-trivial complexities \eqref{eq:example_SU(2)_scaling} and \eqref{eq:complexity_non-trivial_SL2}, respectively.

\section{Relation to gravitational actions and Liouville theory}
\label{sec:Gravity and Liouville theory}
We now proceed to explain the connection between these complexity functionals and Liouville theory.
The coadjoint orbit action of the Virasoro group also arises as an action for the asymptotic degrees of freedom of gravity with asymptotically AdS$_3$ boundary conditions \cite{Barnich,Cotler:2018zff,Mertens:2018fds} (see \cite{Banados:1998gg,Nakatsu:1999wt,Garbarz:2014kaa,Barnich:2014zoa} for earlier work in this direction).
In the same vein, Liouville theory has been derived as the combined action arising from the left and right moving Virasoro symmetries of the asymptotic degrees of freedom of AdS$_3$ spaces \cite{Coussaert:1995zp}.
Moreover, a direct connection between Liouville theory and the sum of coadjoint orbit actions for two copies of the Virasoro group has been derived in \cite{NavarroSalas:1999sr}.
Liouville theory has also appeared in connection with the complexity proposal from path integral optimization of \cite{Caputa:2017urj}.
We will now review the derivation of the above equivalence statements and provide details how the path integral optimization approach of \cite{Caputa:2017urj} is related to the complexity proposal of \cite{Caputa:2018kdj}.

The starting point of the derivation of actions for the asymptotic degrees of freedom of 3 dimensional gravity theories is the Chern-Simons formulation in which the gravity action takes the form
\begin{equation}
  S = \frac{1}{64 \pi G_N} \int (I[A]-I[\bar A])
  \label{eq:Chern Simons action}
\end{equation}
with
\begin{equation}
  I[A] = \Tr \left(A \wedge dA + \frac{2}{3} A \wedge A \wedge A \right).
\end{equation}
For asymptotically AdS$_3$ spaces, $A$ is a $SL(2,\RR)$ valued connection.
We use the conventions
\begin{equation}
  \begin{aligned}
    J_0 = \bar J_0 = \frac{1}{2} \left(
      \begin{array}{cc}
        0 & -1\\
        1 & 0
      \end{array}\right),\\
    J_1 = -\bar J_1 = \frac{1}{2} \left(
      \begin{array}{cc}
        0 & 1\\
        1 & 0
      \end{array}\right),\\
    J_2 = \bar J_2 = \frac{1}{2} \left(
      \begin{array}{cc}
        1 & 0\\
        0 & -1
      \end{array}\right)
  \end{aligned}
\end{equation}
for the $SL(2,\RR)$ generators.
To derive the action of the asymptotic dynamics from \eqref{eq:Chern Simons action}, one imposes asymptotically AdS boundary conditions \cite{Brown:1986nw} which read in terms of the $SL(2,\RR)$ connections \cite{Coussaert:1995zp,Cotler:2018zff}
\begin{equation}
  \begin{aligned}
    A = \left(\begin{array}{cc}
                \frac{dr}{2r} + \CO\left(\frac{1}{r^2}\right) & \CO\left(\frac{1}{r}\right)\\
                rd\sigma^+ + \CO\left(\frac{1}{r}\right) & -\frac{dr}{2r} + \CO\left(\frac{1}{r^2}\right)
              \end{array}\right),\\
    \bar A = \left(\begin{array}{cc}
                     -\frac{dr}{2r} + \CO\left(\frac{1}{r^2}\right) & -rd\sigma^- + \CO\left(\frac{1}{r}\right)\\
                     \CO\left(\frac{1}{r}\right) & \frac{dr}{2r} + \CO\left(\frac{1}{r^2}\right)
                   \end{array}\right),
  \end{aligned}
  \label{eq:asymptotically AdS boundary conditions}
\end{equation}
where $r$ is the radial direction of the bulk and $\sigma^\pm = \sigma \pm t$ with $\sigma$ being the angular and $t$ the time direction.
Moreover, the $SL(2,\RR)$ connections decompose as $A = g^{-1} dg$ and $\bar A = \bar g^{-1} d \bar g$.

The first step of the derivation of both the Liouville and coadjoint orbit actions from the Chern-Simons theory on AdS$_3$ consists of showing that the Chern-Simons theory reduces to two copies of the chiral Wess-Zumino-Witten model.
This follows from a straightforward calculation subject to imposing the boundary conditions $A_- \sim \bar A_+ \sim \CO(1/r)$  from \eqref{eq:asymptotically AdS boundary conditions} \cite{Coussaert:1995zp,Cotler:2018zff}, which gives
\begin{equation}
  S = S_+[g] + S_-[\bar g]
\end{equation}
with
\begin{equation}
  S_\pm[g] = \frac{k}{2\pi} \int_{\partial M} d\sigma dt \Tr((\partial_\sigma g^{-1})(\partial_\pm g)) \mp \Gamma[g] \, .
\end{equation}
Here, $M$ is the bulk manifold in question and $\Gamma[g]$ the topological term of the WZW model.
Note that in order to obtain a variational principle consistent with the AdS$_3$ boundary conditions, a boundary term
\begin{equation}
  S \to S -\frac{k}{4\pi}\int_{\partial M} d\sigma dt (\Tr(A_\sigma^2) + \Tr(\bar A_\sigma^2))
  \label{eq:boundary term CS action}
\end{equation}
has to be added to the action \eqref{eq:Chern Simons action} \cite{Coussaert:1995zp,Cotler:2018zff}.

To obtain the Liouville action, in the second step the two chiral WZW models are combined into a single non-chiral WZW model \cite{Coussaert:1995zp} with action
\begin{equation}
  S = -\frac{k}{\pi} \int_{\partial M} d\sigma dt \,\Tr((\partial_+ \tilde g^{-1})(\partial_- \tilde g)) - \Gamma[\tilde g],
\end{equation}
where $\tilde g = g^{-1}\bar g$.
Then, $\tilde g$ is parametrized in a Gauss decomposition,
\begin{equation}
  \tilde g =
  \left(\begin{array}{cc}
          1 & X\\
          0 & 1
        \end{array}\right)
  \left(\begin{array}{cc}
          e^{\phi/2} & 0\\
          0 & e^{-\phi/2}
        \end{array}\right)
  \left(\begin{array}{cc}
          1 & 0\\
          Y & 1
        \end{array}\right).
  \label{eq:gauss decomposition ordinary WZW}
\end{equation}
Imposing the remaining boundary conditions \eqref{eq:asymptotically AdS boundary conditions} gives rise to the Liouville action\footnote{Strictly speaking, the addition of  a further  boundary term is required to ensure a well-defined variational principle \cite{Coussaert:1995zp}.
  However, upon imposing the AdS$_3$ boundary conditions, this boundary term gives a vanishing contribution to the on-shell value of the action, hence we can neglect it for the purpose of the following discussion.  }
\begin{equation}
  S = \frac{c}{24\pi} \int_{\partial M} d\sigma dt \left(\frac{1}{2} \partial_+ \phi \partial_- \phi + 2e^\phi\right),
\end{equation}
where $c = 6k = 3/2G_N$.

The authors of \cite{Cotler:2018zff}, on the other hand, proceed differently by directly writing $g$ and $\bar g$ in a different Gauss decomposition,
\begin{equation}
  \begin{aligned}
  g = 
  \left(\begin{array}{cc}
          1 & 0\\
          G & 1
        \end{array}\right)
  \left(\begin{array}{cc}
          \lambda & 0\\
          0 & 1/\lambda
        \end{array}\right)
  \left(\begin{array}{cc}
          1 & \psi\\
          0 & 1
        \end{array}\right),\\
  \bar g = 
  \left(\begin{array}{cc}
          1 & -\bar G\\
          0 & 1
        \end{array}\right)
  \left(\begin{array}{cc}
          1/\bar\lambda & 0\\
          0 & \bar\lambda
        \end{array}\right)
  \left(\begin{array}{cc}
          1 & 0\\
          \bar\psi & 1
        \end{array}\right).
  \end{aligned}
  \label{eq:gauss decomposition chiral WZW}
\end{equation}
Imposing the remaining AdS boundary conditions of \eqref{eq:asymptotically AdS boundary conditions} gives
\begin{equation}
  S_+[F] = -\frac{c}{24\pi} \int_{\partial M} d\sigma dt \left(\frac{(\partial_\sigma^2 F)(\partial_+\partial_\sigma F)}{(\partial_\sigma F)^2} - (\partial_\sigma F)(\partial_+ F)\right),
  \label{eq:coadjoint Virasoro action from gravity in vacuum}
\end{equation}
where $G|_{\partial M} = \tan(F/2)$. $S_-[\bar g]$ gives the same contribution up to interchanging $F \to \bar F$ and $\partial_+ \to \partial_-$.
If $F,\bar F$ depend only on $\sigma^+,\sigma^-$ respectively, then we can express the partial derivatives $\partial_+,\partial_-$ w.r.t.~$\sigma^+,\sigma^-$ through partial derivatives $\partial_t$ w.r.t.~$t$.
In doing so, we identify \eqref{eq:coadjoint Virasoro action from gravity in vacuum} with the coadjoint action of the Virasoro group for the vacuum orbit $h=0$ \eqref{eq:geometric_action_Vir}.

The condition that $G,\bar G$ are functions of $\sigma^+,\sigma^-$ resp.~represents an additional constraint that cannot be derived from the equation of motion of $G,\bar G$.
Only for diffeomorphisms $F,\bar F$ that satisfy this constraint, the coadjoint orbit actions from the gravity theory and the coadjoint orbit actions as a complexity functional match\footnote{Note that this can be circumvented by neglecting the boundary term \eqref{eq:boundary term CS action}, in which case one directly obtains \eqref{eq:geometric_action_Vir} \cite{Cotler:2018zff}. However, then the variational principle is no longer well-defined.}.
The constraint is also necessary to derive the equivalence to the Liouville action, as we will see later on.

\medskip
To derive the relation between $F,\bar F$ of the coadjoint orbit action and $\phi$ from the Liouville action, we simply insert the Gauss decompositions \eqref{eq:gauss decomposition ordinary WZW} and \eqref{eq:gauss decomposition chiral WZW} into $\tilde g = g^{-1}\bar g$ to obtain
\begin{equation}
  e^{-\phi/2} = \lambda \bar\lambda (1 + G \bar G).
\end{equation}
Using the fact that the AdS boundary conditions fix $\lambda \sim 1/\sqrt{\partial_\sigma G}$ asymptotically, we obtain
\begin{equation}
  \phi \sim \log\left(\frac{(\partial_\sigma G)(\partial_\sigma \bar G)}{2(1 + G \bar G)^2}\right)
  \label{eq:general solution Liouville eom}
\end{equation}
up to a constant that can be absorbed by a redefinition of $\phi$.
Following \cite{Caputa:2017urj}, we set this constant to zero\footnote{A non-zero value of this constant is equivalent to considering a non-trivial prefactor $\mu \neq 1$ of the $e^\phi$ term in the Liouville action.}.
If $G,\bar G$ depend only on $\sigma^+,\sigma^-$, then this is precisely the general form of the solution of the Liouville equation of motion.

In the path integral optimization approach to complexity of \cite{Caputa:2017urj}, the functions $G,\bar G$ determine the background metric for the path integration and thus are the degrees of freedom which are minimized to obtain the complexity.
Hence, the relation \eqref{eq:general solution Liouville eom} yields a direct mapping between the degrees of freedom that parametrize the complexity functionals of \cite{Caputa:2017urj} and \cite{Caputa:2018kdj}.
There is a small subtlety in this mapping in that the authors of \cite{Caputa:2017urj} use coordinates $(z,x)$ on the plane where $x \in (-\infty,\infty)$, while \cite{Caputa:2018kdj,Cotler:2018zff} work in cylindrical coordinates $(t,\sigma)$ with a periodic coordinate $\sigma \in [0,2\pi)$.
The transformation between these coordinate systems is the same as the transformation between Poincaré patch coordinates and global AdS coordinates near the AdS boundary,
\begin{equation}
  z = \frac{\sin t}{\cos t - \cos \sigma} ~~,~~ x = \frac{\sin \sigma}{\cos t - \cos \sigma}.
  \label{eq:coordinate trafo cylinder plane}
\end{equation}
Using this we may find explicit expressions for the diffeomorphisms $F,\bar F$ for the solutions of the Liouville equation considered in \cite{Caputa:2017urj}.
The first such solution is given by $\bar G = -(x + z)$, $G = 1/(x-z)$.
In \cite{Caputa:2017urj}, the corresponding background metric of the path integral is a time slice of pure AdS$_3$ and the complexity is obtained as the volume of this time slice.
Using \eqref{eq:coordinate trafo cylinder plane} and the definition $G = \tan(F/2)$, $\bar G = \tan (\bar F/2)$, we obtain
\begin{equation}
  F(\sigma,t) = \sigma + t ~~,~~ \bar F(\sigma,t) = \sigma - t + \pi.
  \label{eq:pure AdS3 diffeo from path integral complexity}
\end{equation}
As needed for a consistent diffeomorphism, $F(\sigma + 2\pi,t) = F(\sigma,t) + 2\pi$ and the same for $\bar F$.
Furthermore, these diffeomorphisms are solutions of the equations of motion \eqref{eq:eom_Virasoro} thus they indeed induce optimal transformations leading us from the target to the reference state\footnote{Note that the diffeomorphisms \eqref{eq:pure AdS3 diffeo from path integral complexity} are not only solutions to the equations of motion of the complexity functional \eqref{eq:complexity_Virasoro_with_central_term} including the central extension contribution, but are also solutions to the equations of motion of the functional \eqref{eq:complexity_CM_Vir} without this contribution. In fact, the on-shell value of both complexity functionals is the same for these diffeomorphisms. Thus, in this case, these two complexity measures agree.}.

Inserting \eqref{eq:pure AdS3 diffeo from path integral complexity} into \eqref{eq:coadjoint Virasoro action from gravity in vacuum}, only the second part $-(\partial_\sigma F)(\partial_+ F)$ contributes.
Together with the contribution from $\bar F$, we obtain a complexity of
\begin{equation}
  \CC = S_+[F] + S_-[\bar F] = - \frac{c}{24\pi} \int_0^{T}dt \int_0^{2\pi}d\sigma (-2) = \frac{c}{6} T.
  \label{eq:Liouville complexity h=0}
\end{equation}
The complexity increases linearly in the time $T$ for which the conformal transformation arising from the diffeomorphism acts on the vacuum state.
Identifying $T$ with the inverse UV cutoff (i.e.~letting the conformal transformation act for an infinitely long time), the complexity is proportional to the volume of a time slice of pure AdS$_3$. Thus we reproduce the result of \cite{Caputa:2017urj}.
Such an example for which the Schwarzian term in the coadjoint action vanishes, was also considered in \cite{Caputa:2018kdj}.
Here we arrived at it from a different angle by starting from the equality of the proposals of \cite{Caputa:2018kdj} and \cite{Caputa:2017urj}.

Note, however, that the conformal transformation acting on the reference state is trivial in the sense that it involves only the $L_0$ generator of which the vacuum is an eigenstate.
Hence, the gate acts only by changing the phase of the vacuum state, i.e.~it transforms
\begin{equation}
  \ket 0 \to e^{i c T/6} \ket 0.
\end{equation}
Since this phase change is a gauge symmetry of the problem, the complexity \eqref{eq:Liouville complexity h=0} derived above is not a measureable quantity.

The same applies to a primary state with arbitrary weight $h = \bar h < c/24$ dual to a conical defect in the bulk.
In this case, the action on the corresponding Virasoro orbit takes the form \cite{Cotler:2018zff}
\begin{equation}
  S_+[F] = -\frac{c}{24\pi} \int_{\partial M} d\sigma dt \left(\frac{(\partial_\sigma^2 F)(\partial_+\partial_\sigma F)}{(\partial_\sigma F)^2} - a^2(\partial_\sigma F)(\partial_+ F)\right),
  \label{eq:coadjoint Virasoro action from gravity in pure state} 
\end{equation}
where $a^2 = 1 - 24 h / c$.
Again, for this example the diffeomorphisms $F$, $\bar F$ which follow from the solution of the Liouville equation considered in \cite{Caputa:2017urj} are given by \eqref{eq:pure AdS3 diffeo from path integral complexity}.
Inserting this into \eqref{eq:coadjoint Virasoro action from gravity in pure state}, we obtain a complexity $\CC = c a^2 T / 6$.
As before, the complexity only counts the change of phase in the transformation $\ket{h,\bar h} \to e^{i c a^2 T/6} \ket{h,\bar h}$.

\section{Complexity and the Euler-Arnold method}
\label{sec:Euler-Arnold complexity}
The results of \secref{sec:Examples} and \ref{sec:Gravity and Liouville theory} show that the complexity functional for the Virasoro group measures only phase changes for most cases (except when $\ket h = \ket 0$), calling into question the viability of geometric actions as complexity measures.
The basic problem is that the optimal path in the group manifold with respect to the complexity functional \eqref{eq:complexity_Virasoro_with_central_term} consists of two parts: An instantaneous transition to the target state which comes for free, followed by a phase change with cost proportional to the phase difference.
This problem leads to the conclusion that the cost function leading to \eqref{eq:complexity_Virasoro_with_central_term} should be modified.

Here, we sketch an alternative proposal for a cost function, which was previously alluded to in an appendix of \cite{Caputa:2018kdj}.
This approach avoids the problem of measuring only  phase changes, since in the case the optimal paths  are obtained by application of gates containing all $L_n$ modes.
In contrast, for the geometric action, the optimal path consists of applications of the $L_0$ mode only, generating phase shifts (apart from the instantaneous jump).

The new approach naturally generalizes Nielsen's method \cite{Nielsen,Nielsen_2,Nielsen_3} to arbitrary gate sets possessing a Lie group structure.
The basic idea is to define some metric on the Lie group of the gate set and use the Euler-Arnold method to derive geodesics with respect to this metric.
As summarized in App.~\ref{app:Euler-Arnold}, the Euler-Arnold method provides a simple way of deriving the geodesic equations on a Lie group by replacing the minimization of the length between two points on the group manifold by minimization of an energy functional via the associated Hamilton equations.
The length of these geodesics will then give a complexity definition that naturally generalizes Nielsen's approach.
In fact, complexity definitions for Gaussian states of free quantum field theories inspired by Nielsen's proposal have already employed geodesic distances on Lie groups (although not on the Virasoro group), see e.g.~\cite{Hackl:2018ptj,Chapman:2018hou}.
Let us elaborate on this construction for the Virasoro group in the following.

In general, a positive definite quadratic form on a Lie algebra defines an associated metric on the corresponding Lie group \cite{Arnold}.
The length of a vector in the Lie algebra can then be defined as the cost of the associated infinitesimal symmetry transformation.
Integrating this cost over time gives the total cost for the unitary transformation from reference to target state.
Note that in contrast to the complexity measures of \secref{sec:complexity_general} and \ref{sec:complexity and geometric actions}, the cost at time $t$ depends explicitly   only on the infinitesimal transformation but not on the state on which the transformation is applied to.
The advantage of this method is that the minimization of the total cost can be done by means of the Euler-Arnold method \cite{Arnold}.
This gives a clear and well-understood method of performing the minimization of the total cost, for which a number of solutions are known.
In fact, for the Virasoro case several metrics on the Virasoro group have already been considered in the mathematical literature \cite{Khesin1,Misiolek,Khesin2}.
We will focus on the simplest of these whose associated Euler-Arnold equation is given by the Korteweg-de Vries (KdV) equation.

We will make use of a representation of the Virasoro algebra common in mathematical physics, however perhaps unfamiliar to an audience used to CFT techniques within physics.
Instead of working with the generators $L_n$ satisfying the algebra
\begin{equation}
  [L_n,L_m] = (n-m)L_{n+m} + \frac{c}{12}n(n^2-1)\delta_{n+m,0},
\end{equation}
we use general algebra elements given by linear combinations $\sum_n X_n L_{-n}$ of the generators.
The $X_n$ coefficients are packaged into a periodic function $X(\sigma) = \sum_n X_n e^{in\sigma}$.
Moreover, there is a number $r$ originating from the central extension, so that in total Virasoro algebra elements are represented by the pair $(X(\sigma),r) \equiv (X,r)$.
Writing the Virasoro algebra in this form, the commutator of two of its elements is given by
\begin{equation}
  [(X,r),(Y,s)] = (YX'-XY',\int d\sigma XY''') \label{XY}
\end{equation}
To define geodesics, we need to define a metric, i.e.~a quadratic form $\langle\cdot,\cdot\rangle_g$ \footnote{The $g$ subscript is not an index, but a reminder that the linear functional is associated to a metric.} quantifying how large Virasoro algebra elements are.
The simplest choice is to use the inner product
\begin{equation}
  \langle (X,r), (Y,s) \rangle_g = \int d\sigma XY + rs,
  \label{eq:Virasoro metric}
\end{equation}
which defines a right-invariant $L^2$ metric on the Virasoro group\footnote{Technically speaking, the quadratic form on the Lie algebra $\fg = T_e M$ defines a metric only at the identity element $e$ of the group.
  But of course, elements of a generic tangent spaces $T_g M$ can be transported to the identity tangent space by left/right translation.
  In this way, the inner product on the Lie algebra extends to a metric on the full group manifold.}.
The length of $(X,r)$ is given by $\sqrt{\langle (X,r), (X,r) \rangle_g}$.
For the application to complexity, $X(t,\sigma)$ is identified with the infinitesimal velocity $\epsilon(t,\sigma) = -\dot F/F'$ (see eq.~\eqref{eq:velocity_Vir}).
Then \eqref{eq:Virasoro metric} determines how expensive the infinitesimal conformal transformation induced by $\epsilon$ is, i.e.~it defines the cost function.
In contrast to the cost function \eqref{eq:1_norm_cost_function} used in \secref{sec:complexity_general} and its modification of \secref{sec:complexity and geometric actions}, this cost function is state independent.
It is obtained by mapping the question `How expensive is  the infinitesimal transformation induced by $\epsilon$?' to the question `How long is the representation vector  of the transformation in the Virasoro algebra?'.

To find the optimal transformation with respect to the metric \eqref{eq:Virasoro metric}, i.e.~the one with the lowest cost, we need to derive the geodesic equation on the group manifold.
Using the Euler-Arnold method \cite{Arnold,Arnold2}, the result is (see \cite{Khesin1} or app.~\ref{app:Euler-Arnold} for a short review)
\begin{equation}
  \dot \epsilon + 3\epsilon \epsilon' - r \epsilon''' = 0 ~,~~~  \dot r = 0 \, ,
  \label{eq:KdV}
\end{equation}
with $\epsilon$ the velocity and $r$ as introduced above \eqref{XY}. 
The first of these two equations is known as the KdV equation.
It describes shallow water waves and is well-known for being an exactly solvable non-linear partial differential equation.

To find the associated cost function in terms of the path in the group manifold $F(t)$, we insert the expression $\epsilon = -\dot F/F'$ for the infinitesimal velocities $\epsilon$ in terms of the group elements $F$ into the metric \eqref{eq:Virasoro metric}.
Moreover, we identify an $r$ such that the variation of $r^2$ with respect to $F$ gives $r\epsilon'''$.
Then, the complexity functional is
\begin{equation}
  \CC = \int dt \CF,
  \label{eq:Euler-Arnold complexity}
\end{equation}
where the cost function $\CF$ in terms of Virasoro group elements $F$ is given by
\begin{equation}
  \CF^2 = \int d\sigma \(\frac{\dot F}{F'}\)^2 + \frac{1}{4} \(\int d\sigma \frac{\dot F}{F'} \(\frac{F''}{F'}\)' \)^2.
  \label{eq:Euler-Arnold cost function}
\end{equation}
Assuming that $r = -\frac{1}{2}\int d\sigma\frac{\dot F}{F'} \(\frac{F''}{F'}\)'$ and $\CF$ are constant in $t$, we recover \eqref{eq:KdV} as the equation of motion for \eqref{eq:Euler-Arnold complexity}.
These assumptions are necessary for the derivation of the KdV equation from the Euler-Arnold method to hold (see app.~\ref{app:Euler-Arnold} for details).
They are equivalent to the requirement that the geodesic in the group manifold is traveled along at constant speed.
Of course, the length of the geodesic in independent of this choice, therefore this assumption
does not  place any  restrictions on the allowed target or reference states.

To see whether \eqref{eq:Euler-Arnold complexity} defines a good complexity functional, we need to find infinitesimal velocities $\epsilon$ that solve eq.~\eqref{eq:KdV} such as to follow a geodesic path.
The simplest solutions of \eqref{eq:KdV} are given by $\epsilon(t,\sigma) = \textit{const}$.
Correspondingly, $F(t,\sigma) = G(\sigma - \epsilon t)$.
We note that this is the same kind of solutions as already encountered in \secref{sec:solutions eom Virasoro} for the equations of motion \eqref{eq:eom_Virasoro} of the geometric action.
In particular, since $\epsilon(t,\sigma) = \textit{const.}$ only the $L_0$ generator acts on the reference state, yielding nothing but a phase change.

However, unlike for the equations of motion \eqref{eq:eom_Virasoro} of the geometric action, there also exist non-trivial solutions of \eqref{eq:KdV}, which in the context of complexity lead to non-trivial unitary transformations $U_f$.
These are termed cnoidal waves \cite{KdV} and are of the form
\begin{equation}
  \epsilon(t,\sigma) = \tilde\epsilon + A \,\mathrm{cn}^2\(k\frac{\sigma-Ct}{2\pi}2K(m);m\),
  \label{eq:cnoidal solution KdV}
\end{equation}
where $\tilde\epsilon,A,C,k,m$ are parameters of the solution, $\mathrm{cn}(x;m)$ is a Jacobi elliptic function and $K(m)$ is the complete elliptic integral of the first kind.
In terms of physical parameters of the wave $\tilde\epsilon$ is the trough elevation, $A$ the amplitude and $C$ the phase speed, while $m$ controls the shape of the wave.
For $m \to 0$, the solution reduces to a sine wave and for $m \to 1$ it becomes solitonic in nature.
The parameters are all dependent on each other, thus together with the wavelength $2\pi/k$ the solution has only two free parameters in total.

Now the main question we are interested at this point is: Does the cnoidal wave solutions of the KdV equation suffer from the same problem as the ones for the geometric action, i.e.~does only a phase contribute to the complexity?
To answer that question, we need to determine which $L_n$ modes act at each time step $t$, i.e.~we need to evaluate the Fourier coefficients of $\epsilon(t,\sigma)$.
For a wave length $\lambda = 2\pi$ ($k = 1$), these are straightforwardly determined from the Lambert series of the Jacobi elliptic function,
\begin{equation}
  \mathrm{cn}\((x-Ct)\frac{K(m)}\pi;m\) = \frac{2\pi}{K(m)\sqrt m} \sum_{n=0}^\infty \frac{\cos((n+1/2)(x-Ct))}{2\cosh(\pi(n+1/2)K(1-m)/K(m))},
\end{equation}
to be
\begin{equation}
  \epsilon_n = \tilde\epsilon \delta_{n,0} + A \frac{\pi^2 e^{-inCt}}{4K^2(m)m}\sum_{r=-\infty}^{\infty}\biggl[ \cosh\(\pi(r+\frac{1}{2})\frac{K(1-m)}{K(m)}\) \cosh\(\pi(n-r-\frac{1}{2})\frac{K(1-m)}{K(m)}\) \biggr]^{-1}.
\end{equation}
For other wavelengths $\lambda = 2\pi/k$ with $k \neq 1$, $\epsilon_{n \notin k\ZZ} = 0$ while $\epsilon_{n \in k\ZZ}$ is the same as for $k=1$.
The coefficients quickly converge to zero for $|n| \to \infty$.
The speed of the convergence is controlled by $m$; for $m = 0$ only $\epsilon_{0,\pm k}$ are non-zero, while higher and higher $L_n$ modes contribute as $m \to 1$.
Thus, clearly more Virasoro generators than just $L_0$ contribute to $U_f$ and the resulting non-trivial time dependent conformal transformation $U_f$ does not just generate a phase.
Furthermore, we can control which $L_n$ generators act on the reference state by varying the $m$ parameter of the solution.

Note that the two solutions ($\epsilon = \textit{const.}$ and $\epsilon$ given by a cnoidal wave) of the KdV equation \eqref{eq:KdV} do not exhaust all possibilities.
In particular, there exist solutions that are superpositions of (interacting) cnoidal waves which form a kind of generalization of the Fourier expansion in the solution space of the KdV equation \cite{Osborne}.
These solutions can have an arbitrary number of parameters, instead of essentially just one (the $m$ parameter) for a single cnoidal wave.

In conclusion, the possible unitary transformations $U_f$ obtainable from solutions of the KdV equation are much less constrained than for the equations of motion of the geometric action.
While for the geometric action we only had $\dot f/f' = \textit{const.}$ for $h>0$, here we find --  for any reference state -- an infinite set of solutions with different coefficients of the $L_n$ generators.
To precisely determine which $U_f$ are possible and thus to find the complexity between given reference and target states, one needs to solve for the group element $F$ in the equation $\epsilon = -\dot F/F'$ for $\epsilon$ given by \eqref{eq:cnoidal solution KdV}.
Finding solutions to this non-linear partial differential equation is in general a difficult problem, which we leave for future work.
Nevertheless, it is clear from the above results that the complexity definition \eqref{eq:Euler-Arnold complexity} has a set of known optimal paths over which we have good control and that -- more importantly -- are non-trivial in the sense that they do not instantaneously jump to the target state\footnote{Note also that unitary transformations $U_f$ giving only a phase change can still lead to a non-vanishing complexity \eqref{eq:Euler-Arnold complexity}.
  Therefore, the addition of a boundary term to cancel the phase contribution is necessary to get a well-defined complexity functional from \eqref{eq:Euler-Arnold complexity}.}.

\section{Discussion and conclusion}\label{sec:discuss}
In this paper, we set out to increase our understanding of the notion of complexity for conformal field theories proposed in \cite{Caputa:2018kdj}, which proposes an equivalence  between geometric actions of the Virasoro group and 2d CFTs. Our work revolved around three main points, each of which we now comment on separately.

\subsubsection*{Relation between complexity and geometric actions}

First, we found that the same relation between geometric actions and complexity functionals obtained by the means described in \cite{Caputa:2018kdj} for the Virasoro group also exist for Kac-Moody groups: The geometric actions and complexity functionals match up to terms arising from the central extension of the corresponding group. We then added terms involving the central extension  to the original proposal, such that complexity and geometric action agree exactly.  This was accomplished by considering an additional path through the real numbers determined entirely by the symmetry transformations applied. The additional path then gives an extra contribution to the cost function that coincides with the central term appearing in the geometric action. Therefore, with this generalized cost function, the complexity functionals for the Kac-Moody and Virasoro group are given by their geometric orbit action. This implies an inherent relation between geometric actions and CFT complexity. This connection can be understood in the context of geometric quantization \cite{Kirillov}. As discussed in sec.~\ref{sec:Gauge invariance}, the geometric action is defined on orbits isomorphic to $\hat{G}/H$, where $\hat{G}$ is the centrally extended group and $H$ the orbit stabilizer. For the Virasoro group, we have considered the orbits $\mathrm{Diff}(S^1)/SL(2,R)$ and $\mathrm{Diff}(S^1)/S^1$. These orbits have a Kähler structure if holomorphic and antiholomorphic sectors are combined with the Kähler form corresponding to a symplectic form with one holomorphic and one antiholomorphic index \cite{Witten}. Orbits of this type may be quantized. The Hilbert space then is the space of sections of holomorphic Hermitian line bundles and furnishes a unitary representation of the corresponding group. For instance, for the Virasoro orbit $\mathrm{Diff}(S^1)/S^1$, the Hilbert space is formed by irreducible unitary Verma modules, whereas from the orbit $\mathrm{Diff}(S^1)/SL(2,R)$ we obtain degenerate unitary representations.  

On the other hand, for Kac-Moody groups $\widehat{LG}$ based on compact semisimple Lie groups, the orbits are isomorphic to $\widehat{LG}/T$, where $T$ is from the maximal torus \cite{pressley1988loop}. Similar to the Virasoro case, unitary representations are obtained by means of geometric quantization \cite{Kirillov}. 

These results may be interpreted as follows. The relation between geometric actions and unitary representations implies that after quantization, the complexity functionals \eqref{eq:complexity_Virasoro_with_central_term} and \eqref{eq:complexity_Kac-Moody} yield a Hilbert space that contains the possible target states. Equivalently, if we restrict the allowed gates to symmetry transformations, the possible target states belong to the Hilbert space obtained by geometric quantization from the geometric action. Since the equivalence of the geometric action and the complexity is exact, the complexity functional encodes the possible target states.

In general, it is an interesting question to understand how geometric actions are connected to complexity for symmetry groups.
To this end, in the AdS/CFT context it may be useful to examine the simpler case of theories dual to two dimensional Jackiw-Teitelboim gravity.
In this system, the Schwarzian theory \cite{Maldacena:2016hyu}, based on coadjoint orbits of one copy of the Virasoro group, appears at the boundary.
A complexity definition for the Schwarzian theory was recently given in \cite{Lin:2019kpf}\footnote{We would like to thank Pawel Caputa for pointing this out.}, based on \cite{Brown:2016wib}. An important direction for further work is to
determine the relation between this complexity definition and the geometric action on the coadjoint orbit on which the Schwarzian theory is defined.
\medskip

\subsubsection*{Phase changes and gauge invariance}

Our second point concerns the gauge invariance of the geometric action under the stabilizer group. As discussed in sec.~\ref{sec:Gauge invariance}, the geometric action is invariant only up to total derivatives. These, however, are essential when computing the complexity, which we have identified with the on-shell value of the geometric action. The complexity functionals obtained from generalizing the proposal \cite{Caputa:2018kdj} yield geometric actions without any other contributions and are thus not invariant under the stabilizer group. As we have demonstrated for some examples in sec.~\ref{sec:Examples}, this leads to inconsistencies in the computed complexities, which manifest themselves in different cost for identical transformations.

Furthermore, we obtain non-vanishing cost for target states that only differ by a phase from their reference state. This is explained as follows. The optimal path for the cost function considered consists of two parts. At computation time $t=0$ the transformation immediately jumps to the target state times a phase factor. At later times, only this phase factor changes. For reference states $\ket h$ with $h>0$ and any arbitrary target state, the complexity measures only the phase difference between $U_{f(0)}\ket h$ and $U_{f(T)}\ket h$.

It is possible to cure this problem by adding suitable boundary terms to the complexity action to cancel the contribution of the phase.
In fact, such boundary terms arise naturally in the context of Virasoro Berry phases (see app.~\ref{app:Berry_phases}).
By adding such boundary terms, the complexity between states differing by a phase can be made to vanish.
Then, the complexity measure reduces to a special case of a Berry phase in a coadjoint orbit of the Virasoro group.
However as we have seen in \secref{sec:Examples}, for most reference states (for those with conformal weight greater than zero) the only contribution to the complexity comes from the phase difference.
Thus, the complexity always vanishes for these reference states after the addition of boundary terms. 
Moreover, the equality between the complexity and Liouville or gravity actions no longer holds exactly when adding a boundary term.

The fact that the complexity assigns zero cost to the instantaneous jump relating states at a finite distance in the Hilbert space is surprising also due to the fact that the geometric action is closely related to the canonical distance measure on the Hilbert space induced from the inner product. This distance measure defines a two-norm cost function on the coadjoint orbit, $\CF^2 = \bra{\psi_R} U^\dagger(t)Q(t)Q^\dagger(t)U(t) \ket{\psi_R}$. In \cite{Caputa:2018kdj} it was shown that the leading term of this cost function reduces to the one-norm cost function \eqref{eq:1_norm_cost_function} in the large $c$ limit. Since the corresponding complexity functional is equal to the geometric action up the addition of the central extension term, one would expect the canonical Hilbert space distance measure between reference and target state to give a lower bound for the complexity\footnote{We would like to thank Javier Magán for a discussion about this issue.}. However, we observe that the complexity vanishes even for states which are a finite distance away from each other in the Hilbert space. There are two possible explanations for this discrepancy. Either the central extension term contributes negatively to cancel out the contribution from the Hilbert space distance, or the neglected subleading terms of the two-norm cost function must be taken into account. These subleading terms were recently  studied as a complexity measure in its own right in \cite{Flory:2020eot}. 

Furthermore, our results suggest that the geometric action distinguishes phases from the orbit stabilizer from those of the central extension. Once the boundary terms are added, any phase from the orbit stabilizer is assigned vanishing cost. For a generic optimal transformation, however, the target state picks up an additional phase $e^{ic\int dt \beta(t)}$ from the central extension with cost $c\beta(t)$. This central term generally cannot be cancelled by boundary terms, with one exception. If the optimal transformation giving rise to the phase from the central extension belongs to the stabilizer, the complexity vanishes if appropriate boundary terms are added. This implies that in this case the central term $c\beta(t)$ becomes a boundary term. Another hint that phases from the stabilizer are treated differently than those from the central extension is provided by the equations of motion. The central term modifies the equations of motion, thus leading to different optimal circuits, whereas the contributions from the orbit stabilizer are boundary terms that do not change the equation of motion. This also suggests that the optimal group transformations arising from the geometric action with central extension are not equal to those of the same action without central extension even when projected onto the coadjoint orbit. 

In contrast to the Virasoro case, for Kac-Moody groups, optimal transformations are given by the product of two matrices valued in the semisimple Lie group that specifies the Kac-Moody group, where the first matrix only depends on the position on the unit circle, while the latter only depends on the time it is applied. Therefore, there exist many optimal transformations that do not belong to the orbit stabilizer and thus lead to a non-trivial target state and complexity. We hence conclude that geometric actions present much better cost measures for Kac-Moody groups than for the Virasoro group. This suggests that viable complexity measures may not be universal, even for similar theories such as the CFTs considered here.

\subsubsection*{Relation to path integral complexity}

Our final point is on the relation between the notion of complexity considered here and in the path integral approach \cite{Caputa:2017urj}, which we have shown to be equivalent.
In particular, we made precise the connection between the diffeomorphisms in the geometric action and the solutions of the Liouville equation of motion occurring in the path integral approach.
This allowed us to identify the unitary transformations and target/reference states which in our approach are dual to these solutions.
In particular, we examined the examples of solutions which were argued in \cite{Caputa:2017urj} to be associated to 3-dimensional pure AdS and conical defects.
We found that the dual transformations for these examples are trivial in the sense that they only change the reference state by a phase.
Similar to the geometric action complexity, the phase change produces non-vanishing cost.
In conclusion, the connection between the complexity \eqref{eq:complexity_Virasoro_with_central_term} and the Liouville action found in \secref{sec:Gravity and Liouville theory} maps the volume of a constant time slice of AdS$_3$ to the phase of the vacuum state in the boundary field theory.

While at first this may seem to indicate that the complexity functional of \cite{Caputa:2017urj} 
cannot be a physical quantity, such a conclusion is premature.
In particular, there is no unique way to find a physical interpretation of the path integral optimization approach of \cite{Caputa:2017urj} in terms of gates acting on a reference state.
By a different choice of gate set, reference state and cost function it is still possible to arrive at the same Liouville action as complexity functional, but with differing corresponding transformation from reference to target state.
In fact, such a differing interpretation was already proposed previously in \cite{Camargo:2019isp}.
As in our complexity definition, the gate set proposed in \cite{Camargo:2019isp} consists of exponentials of stress energy tensor components.
However, the authors of \cite{Camargo:2019isp} also include non-unitary operators in their gate set and choose a different, state independent cost function.
Using non-unitary gates naturally avoids the problems of having the complexity measure only a phase shift.
But, of course, it also means that the resulting complexity functional is quite different from the customary definition.
It would certainly be interesting to see if a modification of our complexity functional by including non-unitary gates can solve the problems relating to phase differences encountered in \secref{sec:Examples} and \ref{sec:Gravity and Liouville theory}.
We leave this issue as well as a detailed examination of the relation of our proposal to that of \cite{Camargo:2019isp} to future work.

\subsubsection*{Conclusions about cost functions}

As we discussed extensively in \secref{sec:Examples}, for the Virasoro algebra the complexity counts only the trivial phase changing part of the unitary transformation $U_f$.
This leads to the conclusion that the cost function \eqref{eq:new_cost_function} should be modified.
Which other cost function should one choose?
\medskip

One approach to answering this question is to compare the cost assigned to a conformal transformation with the change of holographic complexity measures under the same conformal transformation acting in the bulk.
The results should agree if the cost function is compatible with holographic complexity.
The corresponding gravity calculation  was performed in \cite{Flory:2018akz,Flory:2019kah}, where the change in complexity under perturbatively small conformal transformations of pure AdS$_3$ was derived for both the ``complexity=volume'' and ``complexity=action'' proposal.
Therefore, the cost of these small conformal transformations applied on the vacuum state is known and the corresponding change in complexity should be reproduced by a cost function derived from field theory arguments.
The results of \cite{Flory:2018akz,Flory:2019kah} show that for a small conformal transformation applied on to the vacuum state the change in complexity is non-vanishing, indicating that the cost function \eqref{eq:new_cost_function} cannot reproduce the holographic results. It was shown in \cite{Belin:2018bpg} that the holographic behavior may be obtained from a notion of complexity defined on the space of Euclidean sources.

More generally, the mapping of the coadjoint orbits to the gravity side of the AdS/CFT correspondence was derived in \cite{Compere:2015knw,Sheikh-Jabbari:2016unm}.
For constant orbit representatives, i.e.~highest weight reference states $\ket h$, the orbits may be classified in three classes according to the value of $h$.
There is one exceptional orbit with $h=0$ for which the dual geometry of the hightest weight state $\ket 0$ is pure AdS$_3$.
The highest weight states of elliptic orbits with $h - c/24 < 0$ are dual to conical defects, while those of the hyperbolic orbits with $h - c/24 > 0$ are dual to BTZ black holes.
Generic orbit elements correspond to excitations of these geometries created by the application of conformal transformations, with the same causal and horizon structure\footnote{Entanglement entropy in such geometries has been investigated in  \cite{Sheikh-Jabbari:2016znt,Ageev:2019fjf}.}.
For example, a generic element of an elliptic orbit corresponds to a particle in AdS$_3$ created by the application of a descendent field to the vacuum state.
Thus, it is known how the bulk geometry changes under a general conformal transformation acting on any reference state $\ket h$.
A field theory cost function must reproduce the corresponding change in the``complexity=volume'' or ``complexity=action'' proposals to match  holographic complexity.
Of course, to fully reproduce holographic complexity proposals also the gate set has to be enlarged.
Conformal transformations alone do not form a set of universal gates since they only connect states in the same Verma module. Nevertheless, in view of the AdS/CFT realization of complexity proposals, a promising new starting point will be to combine the results of the present paper with the results of  \cite{Compere:2015knw,Sheikh-Jabbari:2016unm}. A relation between Virasoro coadjoint orbits and 2d gravity has also been observed in \cite{Mandal:2017thl}.
\medskip

A further approach is to develop criteria to be satisfied by cost functions if they are to yield physically sensible complexity definitions.
\cite{Bueno:2019ajd} examined one such criterion, namely whether the given cost function provides a lower bound for the complexity associated to a discrete gate set\footnote{Cost functions satisfying this criterion were termed $\CG$-bounding in \cite{Nielsen}.}.
Interestingly, it was argued that a good lower bound is achieved by the cost function used by Caputa and Magán \cite{Caputa:2018kdj}, given as eq.~\eqref{eq:1_norm_cost_function} in this paper.
Our results indicate that the criterion of \cite{Bueno:2019ajd} alone is not enough to find a viable cost function.
In particular, for any system possible gauge invariances need to be taken into account when defining cost functions.
The complexity should not assign any cost to gates that give only a non-measureable phase change, a feature which as we have seen is not automatically fulfilled for the cost function \eqref{eq:1_norm_cost_function}.
Note, however, that we do not claim that eq.~\eqref{eq:1_norm_cost_function} suffers from the main problem discovered for geometric actions as complexity functionals -- the fact that geometric actions count only phase changes in all cases but for the vacuum state as reference state.
The equation of motion following from the complexity functional \eqref{eq:complexity_CM_Vir} corresponding to \eqref{eq:1_norm_cost_function} takes on a rather complicated form,
\begin{equation}
  2b\(\frac{\dot f}{f'}\)' + \frac{c}{24}\(
  -2\frac{\dot f'''}{f'} + 4\frac{\dot f'f'''}{f'^2} + 6\frac{\dot f'' f''}{f'^2} - 9\frac{\dot f' f''^2}{f'^3} + \frac{\dot f f''''}{f'^2} + 6\frac{\dot f f''^3}{f'^4} - 6\frac{\dot f f'' f'''}{f'^3}\) \, = 0 \, ,
  \label{eq:eom_complexity_Caputa_Magan}
\end{equation}
for which we did not find non-trivial solutions with $\dot f,f' \neq \textit{const}$.
It is thus conceivable that more involved solutions to this equation lead to non-trivial results for complexity proposals based on
the cost function \eqref{eq:1_norm_cost_function}.
However, the equality to geometric actions on the gravity side and thus also to the Liouville action does not hold exactly for the complexity measure \eqref{eq:complexity_CM_Vir} due to the additional term in the geometric action which we have identified in \secref{sec:complexity and geometric actions} as originating from the central extension\footnote{In \cite{Caputa:2018kdj}, it was claimed that the complexity measure \eqref{eq:complexity_CM_Vir} should be equivalent to the geometric action \eqref{eq:complexity_Virasoro_with_central_term} due to an equality between the equations of motions of both actions.
  We cannot confirm this claim.
  In particular, a standard derivation of the e.o.m.~of \eqref{eq:complexity_CM_Vir} leads us to \eqref{eq:eom_complexity_Caputa_Magan} which is manifestly different from the e.o.m.~\eqref{eq:eom_Virasoro} of the geometric action.
  Moreover, the publication \cite{Aldrovandi:1996sa}, cited to support the claim, only states that the $\int d\sigma dt \dot f/f' (f''/f')'$ part of \eqref{eq:complexity_Virasoro_with_central_term} gives the same equations of motion as $2\int ds dt d\sigma \mu \{f,\sigma\}$,
  when varied w.r.t.~$\mu = \dot f/f'$.
  Note the variation with respect to $\mu(s,t,\sigma)$ under boundary conditions $\mu(0,t,\sigma) = 0$, $\mu(1,t,\sigma) = \dot f(t,\sigma)/f'(t,\sigma)$ and the integral over an additional direction parametrized by $s$, leading to a different contribution than the $\int dt d\sigma \dot f/f' \{f,\sigma\}$ part of \eqref{eq:complexity_CM_Vir}.
}.
\medskip

As explained in \secref{sec:Euler-Arnold complexity}, a further promising direction to explore makes use of the Lie group structure of the gate set in consideration to generalize Nielsen's approach \cite{Nielsen,Nielsen_2,Nielsen_3} to conformal symmetry transformations.
In this case, the cost function is an arbitrary metric on the Virasoro group.
The minimization of the complexity functional can then be easily done using the Euler-Arnold method.
For the simplest possible choice of a right invariant $L^2$ metric on the Virasoro algebra, this yields the Korteweg-de Vries equation as the corresponding equation of motion, as already mentioned in \cite{Caputa:2018kdj}.
This equation has a large number of non-trivial exact solutions which makes analytic calculations possible.
In particular, we may immediately identify non-phase changing unitary transformations among these solutions, thus avoiding the main problem that makes geometric actions unsuitable as complexity functionals.
Other choices of metric lead to different equations of motion well known in the mathematical physics literature, e.g.~the Camassa-Holm or Hunter-Saxon equations \cite{Khesin2}.
Therefore, this method will likely prove fruitful for exploring different cost functions by explicit computations of the corresponding complexity, a task that we leave for future work at this point.
\medskip

\subsection*{Acknowledgements} We are grateful to Souvik Banerjee, Pawel Caputa, Alejandra Castro, Kevin Grosvenor, Javier Magán, René Meyer and Christian Northe for useful discussions.
MG acknowledges financial support from the DFG through Würzburg-Dresden Cluster of Excellence on Complexity and Topology in Quantum Matter - ct.qmat (EXC 2147, project-id 39085490).

\appendix
\section{Relation between complexities and Berry phases}
\label{app:Berry_phases}
Berry phases on orbits of the Virasoro group were considered recently in \cite{Berry_phases}, building on previous work \cite{Mickelsson:1987mx,Bradlyn:2015wsa}.
We give a short review and show that for the Virasoro group the complexity obtained in this publication coincides with the Berry phase up to a boundary term.
For previous work relating the original complexity action of \cite{Caputa:2018kdj} without the central extension piece to Virasoro Berry phases, see \cite{Akal:2019hxa}.

A Berry phase generally arises when a Hamiltonian that depends on an external parameter, for instance a magnetic field, is time-evolved. It is assumed that the parameter adiabatically varies in time and traces out a closed path $\gamma(t)$ in the parameter space. Let $E_n(\gamma(t))$ denote an energy eigenvalue along the path. Note that due to the adiabatic variation of the parameter the level $n$ does not change. An according eigenstate is given by $\ket{\psi_n(\gamma(t))}$. However, the state 
\begin{equation}
\ket{\psi(t)}=e^{iB_n(t)}\ket{\psi_n(\gamma(t))},
\end{equation}
is also an energy eigenstate. Here, $B_n(t)$ denotes the Berry phase induced by the time evolution. It is given by
\begin{equation}
B_n(T)=-\int_0^Tdt\, E_n(\gamma(t))+i\int_{0}^{T}dt\,\bra{\psi_n(\gamma(t))}\partial_t\ket{\psi_n(\gamma(t))}.
\label{eq:ordinary_berry}
\end{equation}
Whereas the 1st term is dynamical, the 2nd is purely geometric as it arises from the dependence of $\ket{\psi_n(\gamma(t))}$ on a point in the parameter space. In particular, the latter may be rewritten in terms of the Berry connection $A_n$,
\begin{equation}
B_{n,\,\mathrm{geo}}=\int_\gamma A_n.
\label{eq:Berry_connection}
\end{equation}
To see the relation to the complexity, we need to generalize the Berry phase to symmetry groups, i.e. the parameter space must become a group manifold such that we consider paths $g(t)$ in the group manifold as we have done for the complexity. We now schematically illustrate how this transition from parameter spaces to group manifolds is accomplished. 

Let $G$ be a connected Lie group with group elements $g$ containing a one-parameter subgroup, which we interpret as time translations in order to make the connection to the "ordinary" Berry phases described above. To introduce a similar dependence of the Hamiltonian on the parameter space, the Hamiltonian must now depend on group elements $g$. This is most easily interpreted as choosing different reference frames, i.e. the Hamiltonian $\hat H$ is equally suitable as the transformed one $U_g\hat HU^{\dagger}_g$. If the path is closed, implying $g(T)=g(0)$, \eqref{eq:ordinary_berry} yields the correct Berry phase on the group manifold. However, the group paths considered in the complexity context are typically not closed. We may put a milder restriction on the path by requiring $g(0)$ and $g(T)$ belong to the same ray, i.e. 
\begin{equation}
U_{g(T)}\ket{\psi}=e^{i\theta}U_{g(0)}\ket{\psi}, 
\end{equation}
where $\theta\in\mathbb{R}$. Here, an important aspect comes into play: it is actually not the path $g(t)$ itself that is relevant for the Berry phase, but its projection on the manifold $G/H$, where $H$ is a subgroup of $G$ corresponding to the stabilizer of $\ket{\psi}$. If the expectation value in \eqref{eq:ordinary_berry} is then evaluated in a highest-weight representation, this manifold is a coadjoint orbit of $G$. This is equivalent to what we have done for the complexity as discussed in the previous subsection. We wrote the complexity functional in terms of a pre-symplectic form on $G$, not on the coadjoint orbit itself. Similarly, the Berry curvature arising from the Berry connection in \eqref{eq:Berry_connection} is not defined on the coadjoint orbit, but rather on $G$ and is equivalent to the pre-symplectic form \eqref{eq:presymplectic_form}. Just as discussed for the complexity, only the projected path on the orbit is relevant. For the Berry phase, this implies only the path on the orbit must be closed. This amounts to requiring that $g^{-1}(0)g(T)\in H$. Furthermore, to ensure gauge invariance under $H$, a boundary term is introduced in \eqref{eq:ordinary_berry}. Generalized to group manifolds, the Berry phase then reads
\begin{equation}
B_{\mathrm{geo}}=\int_g dt\,i\bra{\psi}U^{\dagger}_gdU_g\ket{\psi}-i\mathrm{log}\bra{\psi}U_{g^{-1}(0)g(T)}\ket{\psi}.
\label{eq:berry_group}
\end{equation}
The 1st term is just the Maurer-Cartan form in terms of $G$ in a unitary representation and may be generalized to centrally extended groups by including its central extension. Evaluating the Berry phase for the Virasoro group in terms of the group path $f(t)$ then yields
\begin{equation}
B_{\mathrm{geo}}=-\frac{1}{2\pi}\int_0^T dt\int_0^{2\pi} d\sigma\,\frac{\dot{f}}{f^{\prime}}\left[h-\frac{c}{24}+\frac{c}{24}\(\frac{f^{\prime\prime}}{f^{\prime}}\)^{\prime}\right]+\(h-\frac{c}{24}\)F(0,f(T,0)),
\label{eq:Berry_phase_Vir}
\end{equation} 
where $F$ is the inverse of $f$ and $h>0$ is the conformal weight of the highest weight state $\ket\psi = \ket h$ defining the coadjoint orbit.
Apart from the boundary term $(h-c/24)F(0,f(T,0))$, this is exactly equal to the complexity functional \eqref{eq:complexity_Vir}.
The boundary term arises from the logarithmic term in \eqref{eq:berry_group}.
It ensures that the Berry phase vanishes for transformations $f(t,\sigma)$ belonging solely to the stabilizer $H$ of $\ket\psi$.

Furthermore, all solutions to the equations of motion \eqref{eq:eom_Virasoro} define closed paths in coadjoint orbits, as we will show now.
Remember that the general solution to \eqref{eq:eom_Virasoro} is given by $f(t,\sigma) = g(f_0(t,\sigma))$, where $g$ is an arbitrary diffeomorphism and $f_0$ is given by \eqref{eq:solution eom Virasoro h>0} for $h>0$ and by \eqref{eq:solution eom Virasoro h=0} for $h=0$.
The inverse of such a solution is given by $F(t,\sigma) = F_0(t,G(\sigma))$, where $G,F_0$ are the inverses of $g,f_0$ respectively.
Then it follows that
\begin{equation}
  F(0,f(T,\sigma)) = F_0(0,f_0(T,\sigma)) = \left\{
    \begin{aligned}
      \sigma + \theta ~&,~~ h > 0\\
      2\arctan\(\frac{A\tan(\sigma/2) + B}{C\tan(\sigma/2) + D}\) ~&,~~ h = 0,
    \end{aligned}
  \right.
  \label{eq:Berry phase condition closed path}
\end{equation}
with $AD-BC = 1$.
The quantity on the right hand side is the identity $f(\sigma)=\sigma$ transformed by some element of the stabilizer subgroup $H$ of the orbit (U(1) for $h>0$ and SL(2,R) for $h=0$).
Eq.~\eqref{eq:Berry phase condition closed path} is the explicit form of the condition $g^{-1}(0)g(T) \in H$ encoding the requirement that the path $f(t,\sigma)$ is closed \emph{on the orbit}.
The important thing to notice now is that the solutions of the equations of motion \eqref{eq:solution eom Virasoro h>0}, \eqref{eq:solution eom Virasoro h=0} fulfill \eqref{eq:Berry phase condition closed path}.

Therefore, we conclude that the complexity for all possible paths $f(t,\sigma)$ in the Virasoro group manifold is equivalent to a Berry phase up to a boundary term which for $h>0$ is given by $(h-c/24)F(0,f(T,0))$.
Note that the converse is not true, there exist Berry phases that are not equivalent to a complexity.
These are generated by closed paths $f(t,\sigma)$ which are non optimal, i.e.~paths which do not solve the equations of motion of the geometric action \eqref{eq:complexity_Virasoro_with_central_term}.

\section{Euler-Arnold method for the Virasoro group}
\label{app:Euler-Arnold}
The Euler-Arnold method \cite{Arnold,Arnold2} provides a convenient way of deriving geodesic equations on Lie groups.
To make this publication self contained, we provide a short review of the method in general and its application to the Virasoro group \cite{Khesin1,Misiolek,Khesin2}.


Take some curve $X(t)$ on a Lie algebra and a linear functional $\langle\cdot,\cdot\rangle_g$ defining a metric on the corresponding Lie group.
The length of the curve with respect to this metric is given by
\begin{equation}
  \ell = \int_0^Tdt \sqrt{\langle X, X \rangle_g}.
  \label{eq:length of geodesic}
\end{equation}
Geodesics are the curves with minimal length.
The main simplification of the Euler-Arnold method is due to replacing minimization of $\ell$ by minimization of the ``energy''
\begin{equation}
  E = \frac{1}{2} \int_0^Tdt \langle X, X \rangle_g.
  \label{eq:energy along geodesic}
\end{equation}
The two minimal curves obtained in this way coincide provided that the curve is traversed at constant speed, i.e.~$\langle X, X \rangle_g$ is constant in $t$.
Due to the reparametrization invariance in $t$ enjoyed by the problem this is always a valid choice to make.

To derive the geodesic equation, it is useful to relate the metric $\langle\cdot,\cdot\rangle_g$ to the pairing $\langle \cdot , \cdot \rangle$ between elements of the Lie algebra $\mathfrak{g}$ and its dual space $\mathfrak{g^*}$.
This is achieved by defining the \emph{inertia operator} $A$ mapping the velocity $X \in \mathfrak{g}$ to a corresponding element of $v \in \mathfrak{g^*}$ known as the \emph{intrinsic momentum}.
The inertia operator is defined by
\begin{equation}
  \langle X, Y \rangle_g = \langle AX, Y \rangle.
\end{equation}
It is common to write the Hamiltonian associated to the energy \eqref{eq:energy along geodesic} in terms of the intrinsic momentum,
\begin{equation}
  H = \frac{1}{2} \langle AX, X \rangle = \frac{1}{2} \langle v, A^{-1}v \rangle.
\end{equation}

Then one can show that the geodesic equation for \eqref{eq:length of geodesic} is equivalent to the Euler-Arnold equation on $\fg^*$ \cite{Arnold,Arnold2}\footnote{The form of the Euler-Arnold equation \eqref{eq:Euler-Arnold equation} shown here is also known as Lax pair form or the Lie-Poisson equation. The Euler-Arnold equation formulated on $\fg$ instead of $\fg^*$ is also termed the Euler-Poincaré equation.}
\begin{equation}
  \dot v = -\mathrm{ad}^*_{A^{-1}v}v,
  \label{eq:Euler-Arnold equation}
\end{equation}
where the \emph{coadjoint operator} $\mathrm{ad}^*_X$ is implicitly defined through the relation
\begin{equation}
  \langle \mathrm{ad}^*_Xv,Y \rangle = \langle v,[X,Y] \rangle.
  \label{eq:coadjoint operator}
\end{equation}
To simplify the notation, we present the proof only for the case of matrix groups.
Then $X$ is related to the group elements $g$ specifying the path as $X = \dot g g^{-1}$.
Perturbing the path $g \to g + \delta g$ gives a perturbation $\delta X = \delta \dot g g^{-1}- \dot g \delta g^{-1}$ and
\begin{equation}
  \delta H = \langle AX, \delta X \rangle = \langle v, \delta \dot g g^{-1} - \dot g g^{-1} \delta g g^{-1} \rangle.
\end{equation}
Partially integrating the first term in $t$ and applying the definition \eqref{eq:coadjoint operator} of the coadjoint operator gives
\begin{equation}
  \delta H = - \langle \dot v, \delta g g^{-1} \rangle + \langle v, [\delta g g^{-1}, X] \rangle = -\langle \dot v + \mathrm{ad}^*_{A^{-1}v} v, \delta g g^{-1} \rangle.
\end{equation}
Demanding that $\delta H = 0$ vanishes for arbitrary perturbations leads to \eqref{eq:Euler-Arnold equation}.

Now we specialize to the Virasoro group.
In this case, $X \equiv (X(\sigma)\partial_\sigma,r)$, where $X(\sigma)$ is a periodic function of $\sigma$ and $r$ is a constant belonging to the central extension.
Elements of the dual space are denoted by $v \equiv (v(\sigma) d\sigma^2,c)$ where again $v(\sigma)$ is a periodic function and $c(t)$ a constant.
The natural pairing between elements of $\fg$ and $\fg^*$ is given by
\begin{equation}
  \langle (v d\sigma^2,c), (X\partial_\sigma,r) \rangle = \int d\sigma vX + cr.
\end{equation}
Thus, for the choice of metric \eqref{eq:Virasoro metric}, $A$ simply maps $(X\partial_\sigma,r) \in \fg \to (X d\sigma^2,r) \in \fg^*$.
However, there are of course other possible choices of inertia operator leading to different metrics. (For example, in \cite{Khesin2} $A: (X\partial_\sigma,r) \to ((\alpha X-\beta X'') d\sigma^2,r)$ has been considered.
For $\alpha = \beta = 1$ the resulting geodesic equation is the Camassa-Holm equation $\dot v - \dot v'' + 3 vv' - 2 v'v'' - vv''' - cv''' = 0$.)
The coadjoint operator $\mathrm{ad}^*_X$ for the Virasoro algebra is given by \cite{Khesin2}
\begin{equation}
  \mathrm{ad}^*_{(X\partial_\sigma,r)}(v d\sigma^2,c) = ((2vX' + Xv' - cX''')d\sigma^2, 0).
\end{equation}
Inserting this in \eqref{eq:Euler-Arnold equation} and using the definition of $A$ for the metric \eqref{eq:Virasoro metric} yields the KdV equation \cite{Khesin2}
\begin{equation}
  \dot v + 3vv' - cv''' = 0 ~,~~ \dot c = 0.
\end{equation}
Identifying $c = r$ and $v$ with the infinitesimal velocity $\epsilon$ gives the desired result, eq.~\eqref{eq:KdV}.

Further insight into the Euler-Arnold equation can be obtained by considering the evolution of an arbitrary $v$ under \eqref{eq:Euler-Arnold equation}.
One can show that orbits of the coadjoint representation are \emph{invariant manifolds} under this evolution, i.e.~starting from a $v(0)$ in some coadjoint orbit, we will evolve to a $v(t)$ in the same coadjoint orbit \cite{Arnold}.
For the Virasoro case, this is just the familiar statement that acting with conformal transformations on a state in some Verma module gives another state in the same Verma module.
Thus, the coadjoint orbit, which tells us which reference state we are considering, is implicitly contained in the choice of boundary condition $v(0)$ for the Euler-Arnold equation.
Furthermore, the energy $H = 1/2 \langle v, A^{-1} v \rangle$ is a Hamiltonian function defined on a coadjoint orbit \cite{Arnold}.
The Euler-Arnold equations \eqref{eq:Euler-Arnold equation} are then nothing but the Hamilton equations for this function.

\bibliographystyle{JHEP}
\bibliography{bibliography}

\end{document}